\documentclass[fleqn,usenatbib]{mnras}

\usepackage{ulem}

\usepackage[T1]{fontenc}

\DeclareRobustCommand{\VAN}[3]{#2}
\let\VANthebibliography\thebibliography
\def\thebibliography{\DeclareRobustCommand{\VAN}[3]{##3}\VANthebibliography}

\usepackage{graphicx}	
\usepackage{amsmath}	
\usepackage{amssymb}	
\usepackage{tabularx} 
\usepackage[vlines]{tabularht}
\usepackage{booktabs}
\usepackage{multicol}
\usepackage{multirow}
\usepackage{subcaption}
\usepackage{float}
\usepackage{newtxtext,newtxmath}

\newcommand{\rev}[1]{\textcolor{black}{#1}}
\newcommand{\revtwo}[1]{\textcolor{black}{#1}}
\newcommand{\revthree}[1]{\textcolor{black}{#1}}

\title{The Loneliest Galaxies in the Universe: A GAMA and GalaxyZoo Study on Void Galaxy Morphology}

\author[L. E. Porter et al.]{
Lori E. Porter,$^{1}$\thanks{E-mail: lori.porter@louisville.edu}
Benne W. Holwerda,$^{1}$
Sandor Kruk,$^{2,3}$
Maritza Lara-L\'opez,$^{4}$
Kevin Pimbblet,$^{5}$
\and
Christopher Henry,$^{1}$
Sarah Casura,$^{6}$ and
Lee Kelvin$^{7}$ \\
$^{1}$Department of Physics and Astronomy, University of Louisville, 102 Natural Sciences Building, Louisville, KY 40292, USA\\
$^{2}$Max-Planck-Institut f\"ur extraterrestrische Physik (MPE), Giessenbachstrasse 1, 85748 Garching bei M\"unchen, Germany\\
$^{3}$European Space Agency, ESTEC, Keplerlaan 1, NL-2201 AZ, Noordwijk, the Netherlands\\
$^{4}$Departamento de Física de la Tierra y Astrofísica, Instituto de Física de Partículas y del Cosmos, IPARCOS, Universidad Complutense de Madrid (UCM), 28040 Madrid, Spain \\
$^{5}$ E.A. Milne Centre for Astrophysics, University of Hull, Cottingham Road, Kingston-upon-Hull, Hull HU6 7RX, UK \\
$^{6}$ Hamburger Sternwarte, Universit\"at Hamburg, Gojenbergsweg 112, 21029 Hamburg, Germany\\  
$^{7}$Department of Astrophysical Sciences, Princeton University, 4 Ivy Lane, Princeton, NJ 08544, USA
}

\date{Accepted 2023 April 10. Received 2023 April 10; in original form 2023 February 24}

\pubyear{2023}

\begin{document}
\label{firstpage}
\pagerange{\pageref{firstpage}--\pageref{lastpage}}
\maketitle

\begin{abstract}
The large-scale structure (LSS) of the Universe is comprised of galaxy filaments, tendrils, and voids. The majority of the Universe’s volume is taken up by these voids, which exist as underdense, but not empty, regions. The galaxies found inside \revthree{these} voids are expected to be some of the most isolated objects in the Universe. This study, using the Galaxy and Mass Assembly (GAMA) and Galaxy Zoo \revtwo{surveys}, aims to investigate basic physical properties and morphology of void galaxies versus field (filament and tendril) galaxies. We use void galaxies with stellar masses \revtwo{($\rm M_*$)} of $10^{9.35} M_{\odot} < M_* < 10^{11.25} M_{\odot}$, and this sample is split by identifying \rev{two redshift-limited regions, 0 < z < 0.075,  and, 0.075 < z < 0.15.} To find comparable \revtwo{objects} in the sample of field galaxies from GAMA and Galaxy Zoo, we identify \revthree{`}twins' of void galaxies as field galaxies within $\pm$0.05 dex and $\pm$0.15 dex \rev{of $\rm M_*$ and \revtwo{specific star formation rate}.} \revtwo{We determine the statistical significance of our results using the Kolmogorov-Smirnov (KS) test.} We see that void galaxies, in contrast with field galaxies, seem to be disk-dominated and have predominantly round bulges \revtwo{(with > 50\% of the Galaxy Zoo \revthree{citizen scientists} agreeing that bulges are present).} 
\end{abstract}

\begin{keywords}
galaxies: evolution -- galaxies: spiral -- galaxies: structure -- galaxies: formation
 \end{keywords}



\section{Introduction}

Void galaxies are expected to be some of the most isolated objects in the Universe. However, their standard morphology, and how it compares to galaxies in denser regions of the Universe, remains a topic of debate. Studying galactic morphology, how galaxies are classified, and possible links between physical properties and morphological type is essential to further developing our understanding of galaxy formation and evolution. 

Galaxy environment is arguably one of the most important factors in determining what shape a galaxy takes \citep{Dressler1984, Postman2005, Hambleton2011, Buta2015}. With this dependence on environment, it is not unreasonable to believe that the secluded nature of void galaxies could have a substantial effect on their morphology. With fewer merging galaxies in these underdense regions, a lack of clusters, and less material for accretion, the evolution of these galaxies is highly likely to be driven by internal processes. As a result, void galaxies are optimal natural laboratories for studying how galaxies evolve in isolated environments, which can possibly explain how important morphological features form \citep{Kormendy1979b, Combes&Sanders1981}. While \citet{Hambleton2011} points out weaknesses in relying on broad morphological classifications, it is first important for us to understand these broad categorical distinctions in void galaxy morphology before we can study the finer details.

Understanding the basic morphology of void galaxies (if one exists) provides a gateway to possibly linking other physical properties of void galaxies with environment. There also \revtwo{exist} well-known relations involving morphology, environment, and other galaxy properties, such as the \revtwo{star formation rate (SFR)}-$\rm M_*$-morphology relation \citep{Blanton2003, Kauffmann2003, Wuyts2011, Kelvin2014b}, environment and mass-quenching \citep{Peng2010}, and star formation-morphology \citep{Kennicutt1998, Williams2009, Barro2013, Kelvin2018}. For example, galaxies in denser environments have been found to be redder in color, have lower star formation rates, and be more elliptical, typically caused by the neighboring galaxies and higher incidents of mergers \citep{Dressler1984, Kauffmann2003, Lotz2008, Lotz2011, Peng2010, Bell2012a, Alpaslan2015, Woo2015}. On the other hand, galaxies in the lower-density areas are usually largely dominated by spiral galaxies \citep{Dressler1984}.

\citet{Rojas2004} \revtwo{identify} nearly a thousand void galaxies \revthree{in} the Sloan Digital Sky Survey (SDSS) using a nearest neighbor analysis. \revtwo{They investigate} the \rev{S\'ersic} index in two populations, wall galaxies (non-void galaxies, also known as tendril and filament galaxies) and void galaxies in two distance groups: near and distant. \rev{This results in a total of two statistical tests being conducted: a comparison of the \rev{S\'ersic} index in the nearby void galaxies versus nearby wall galaxies, and distant void galaxies versus distant wall galaxies.} \revtwo{They find} no significance in the Kolmogorov-Smirnov statistics between the nearby groups, but statistical significance in the distant sample. 

These are conflicting results, and result in an inconclusive study in terms of galaxy morphology. However, \citet{Rojas2004} \revtwo{do} determine that void galaxies appear to be bluer in color and fainter than wall galaxies in both the nearby and distant samples.

In addition, other studies agree with the conclusion that void galaxies are expected to have higher specific star formation rates (sSFR) and retain more of a blue color as compared to similar galaxies in more dense environments \citep{Rojas2004, Rojas2005, Hoyle2012, Moorman2015, Penny2015, Moorman2016, Beygu2016, Beygu2017, Florez2021}. However, \citet{Kreckel2014} \revtwo{disagree}, stating that in their sample of 61 void galaxies in the Void Galaxy Survey (VGS), there appeared to be no evidence for bluer colors at a fixed luminosity (although the authors note their small sample size and the need for control of all variables), and that void galaxies have similar gas disks to galaxies in denser environments. The analysis of nine void galaxies from SDSS Data Release 7 (DR7) by \citet{Fraser-McKelvie2016} \rev{ and \citet{Ricciardelli2014} also suggest} that the isolation of void galaxies has no effect on the SFR. \rev{ \citet{Rosas-Guevara2022}, on the other hand, \revtwo{note} that similarities in SFR seem to vary depending on stellar mass ($M_*$).}

\citet{Rojas2005} \revtwo{suggest} that void galaxies have more spirals than their counterparts in denser environments, with \citet{VanDeWeygaert2011} suggesting that they maintain a late-type morphology. In addition, \citet{Beygu2016} \revtwo{find} that void galaxies from the Void Galaxy Survey typically have a lower \rev{S\'ersic} index (n $<$ 2), typically \revtwo{indicative} of more disky galaxies, but concludes that void galaxies do not seem to have a specific type. \rev{\revthree{Conversely,} \citet{Penny2015} \revtwo{find} that void galaxies do not exhibit a different morphology than those in denser environments, in addition to other properties mentioned above.}

\citet{Pustilnik2019}'s analysis of dwarf galaxies in voids \rev{shows} that these galaxies typically have morphologies consistent with irregular (morphologies that are neither elliptical nor spiral) and late-type spiral galaxies, \rev{quantitatively suggesting that 7\% of local void galaxies are early-types, 41.6\% are some type of spirals, and 43.2\% are irregular. The remaining galaxies are either blue compact objects or lenticulars.}

\rev{ \citet{Florez2021} \revtwo{suggest} that void galaxies altogether follow a specific evolutionary path, dependent on the dark matter halo. When investigated at a fixed mass, void galaxies here agree with previous results in that they are bluer, star-forming, and gas-rich, and that these trends persist with morphology as well. The authors note that this is likely due to a galaxy assembly bias, and indeed find that the trends are replicated when galaxy properties are matched to halo properties. }

\rev{Indeed, simulations and theory further bolster the need to investigate correlations between galaxy environment and morphology. \citet{Croton2008} \revtwo{investigate} a population of quenched late-type void galaxies, comparing their luminosity functions to galaxy formation models built from Millenium simulations. \revtwo{Their results suggest} that despite their large-scale environmental differences, galaxies residing in similar dark matter halo masses will retain similar properties.}

\rev{ \citet{Rosas-Guevara2022} \revtwo{provide} a new perspective by using the EAGLE hydrodynamical cosmological simulations to investigate void galaxy properties and their assembly histories. After controlling for the effect of stellar halo mass, \citet{Rosas-Guevara2022} finds that their sample of most isolated void galaxies have the fewest positive gas-phase metallicity gradients present. This finding alludes to the possible association between external processes and feedback events in isolated environments, which implies that these most-isolated galaxies have fewer instances of mergers than their analogues in denser environments.}

Clearly, results and sampling of void galaxies remain diverse across studies and often lead to conflicting results. Therefore, this study aims to remedy this problem by using new data and a variety of perspectives.

\citet{Alpaslan2014} introduces a new spectroscopically complete catalogue of the large-scale structure (LSS) of the Universe called the Galaxy and Mass Assembly (GAMA) Large Scale Structure Catalogue (GLSSC), comprising over forty thousand galaxies. \revtwo{They} identify \revtwo{each galaxy} as belonging to either filaments (the largest structure), tendrils (the second-largest structure, and substructure of filaments), or voids. Because of the introduction of tendrils, in addition to filaments, galaxies can be more accurately grouped according to their environment.

This study introduces the idea of combining the powerful sample created by \citet{Alpaslan2014} with the resources in Galaxy Zoo, to complete an observational analysis on void galaxy morphology. This paper is organized as follows: we begin by reviewing the surveys from which our sample is selected in Section \ref{sec:data}, specifically elaborating on how void galaxies are identified in Section \ref{sec:vg}, and how we selected our analysis sample in Section \ref{sec:sampleselection}. We go over our results from both GAMA and Galaxy Zoo in Section \ref{sec:results}, discuss interpretations in a physical sense and compare with previous literature in Section \ref{sec:discussion}, and finally briefly summarize this study in Section \ref{sec:conclusions}.

\section{Data}
\label{sec:data}

All galaxies are identified from the Galaxy And Mass Assembly (GAMA) survey \citep{Driver2009, Liske2015}. We combine the GAMA Data Release 3 \citep{Baldry2018} 
and the KiDS \citep{DeJong2013, DeJong2015, DeJong2017, Kuijken2019} imaging, with MAGPHYS computing the stellar mass and specific star formation rate (sSFR) \rev{utilized} in this study \citep{DaCunha2008}. In addition, we use the GLSSC from \citet{Alpaslan2014} to identify void galaxies, and morphology voting is from the Galaxy Zoo GAMA-KiDS project. 

\subsection{GAMA}
\label{sec:GAMA} 

GAMA is a highly complete (>98\% to r < 19.8 mag) spectroscopic and multiwavelength imaging survey conducted with the intent to investigate large-scale structure (LSS) in the local Universe \rev{(z < 0.6)} on kpc to Mpc scales \citep{Driver2009, Driver2011, Driver2022, Baldry2018}. The survey now consists of \revtwo{five regions}, three of which are equatorial regions of 5 degrees in declination and 12 degrees in right ascension, \rev{covering a total of nearly \revthree{250,000} galaxies. Additional photometric data was collected on each galaxy in 20+ bands at multiple wavelengths \citep{Liske2015, Driver2016, Baldry2018, Driver2022}. This specific study uses GAMA Data Release \revthree{4}, detailed in \citet{Driver2022}, where the galaxies' S\'ersic indices and effecive radii are computed by \citet{Kelvin2012} \revtwo{in SersicPhotometryv09}.  With such a large and complete sample of high resolution data, we are well-equipped to study the selected population of galaxies.}

\subsubsection{MAGPHYS}

\revthree{As part of GAMA, the MAGPHYS v06} spectral energy distribution (SED) fit data products \citep{DaCunha2008, Driver2009, Driver2011, DaCunha2015} to calculate physical properties such as specific star formation rate (sSFR), redshift, and stellar mass,  accounting for the emission from stellar populations, and both dust attenuation and emission. For further details on MAGPHYS, we direct the reader to \citet{DaCunha2008} and \citet{DaCunha2015}. \rev{This allows  us to further select field galaxies for comparison that are effectively identical to void galaxies in terms of star formation, as described in Section \ref{sec:sampleselection}.}

\subsubsection{Void Galaxies}
\label{sec:vg}

Void galaxies are defined by \citet{Alpaslan2014} as a galaxy that is at a minimum 4.56h$^{-1}$ Mpc from the nearest tendril galaxy, which are a minimum of 4.12h$^{-1}$ Mpc from filaments. This survey samples galaxies from various stellar mass groups, which allows for trends caused by environment to be more prevalent than trends in galaxies caused by mass. \citet{Alpaslan2014} and \citet{Alpaslan2015} then use data from \citet{Pan2012}, to identify a new sample of void galaxies that are truly isolated, and prove that many galaxies previously identified as voids may actually be tendril galaxies. As a result, the galaxies identified \revtwo{by FilamentFindingv02} in \citet{Alpaslan2014} are expected to truly be some of the most isolated objects in the Universe. These parameters and the high resolution of GAMA allow us to be \revtwo{more confident} that these void galaxies are truly isolated. For ease, we will now refer to any galaxy that is \textit{not} a void galaxy (i.e, a tendril or filament galaxy) as a field galaxy. 

\begin{figure*}
	\includegraphics[scale=0.6]{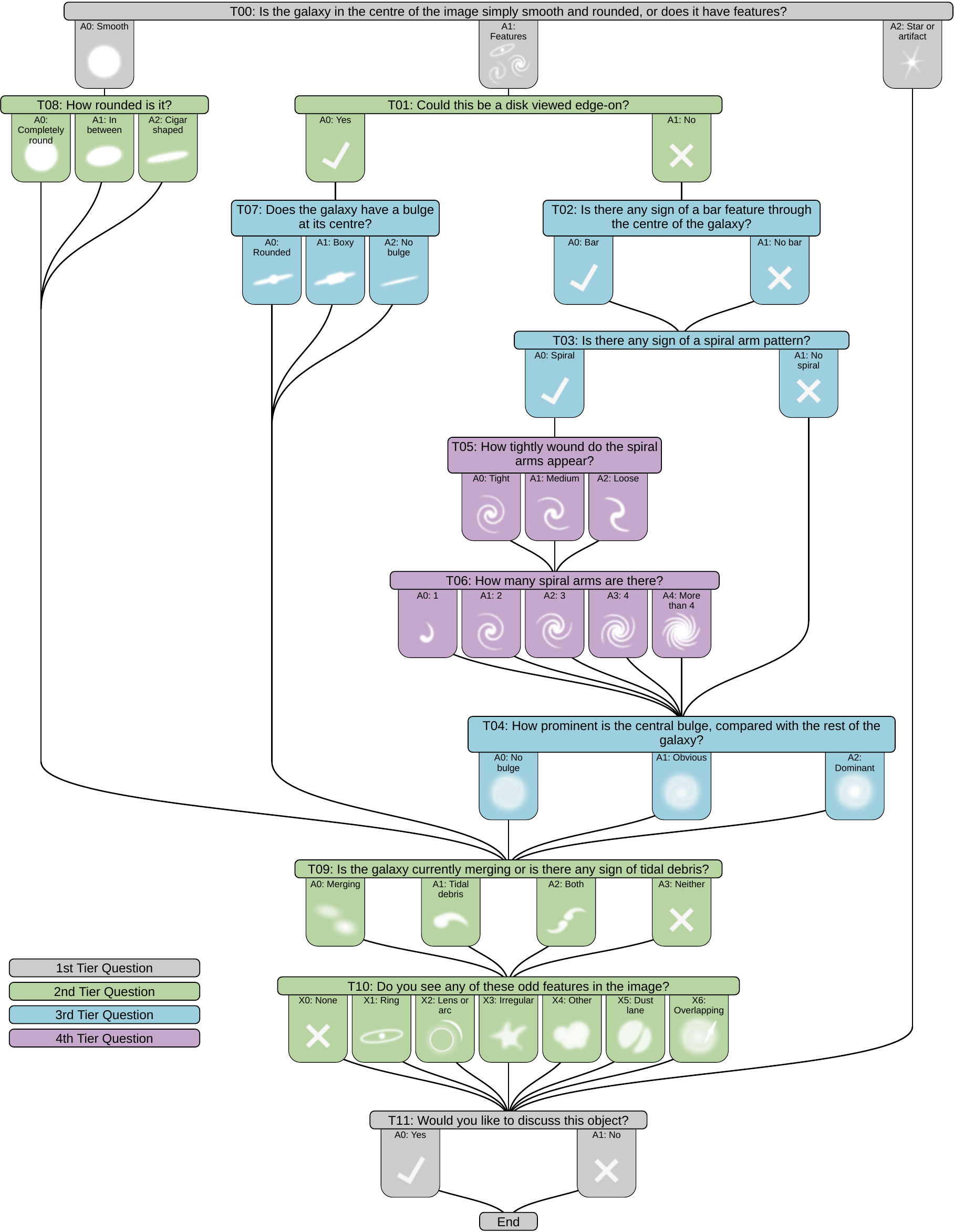}
    \caption{ The Galaxy Zoo decision tree for the GAMA-KiDS GZ survey. Participants begin at the top of the tree with the first question, color-coded by the key visible in the bottom-left, and move their way throughout the tree based on their answers to each question. This study focuses on questions T00, T02, T03, T04, T07, and T09. \revtwo{Note that later in this study for question T09, to avoid redundancy, we simply combine the answers for "merging", "tidal debris", and "both", effectively limiting the possible answers to T09 to "Yes" or "No."}}
    \label{fig:gz_tree}
\end{figure*}

\subsection{GAMA-KiDS Galaxy Zoo}
\label{sec:GZ}

Our analysis on void galaxy morphology is largely based on the GAMA-KiDS Galaxy Zoo survey (Kelvin et al. in prep.). 49,851 galaxies are selected from GAMA equatorial fields with a maximum redshift of z=0.15 for use in morphological classification, with questions in the survey following the structure shown in Figure \ref{fig:gz_tree}. Comprehensive voting fractions are then evaluated, with each voting fraction representing the portion of the population that vote for a specific component's presence (or lack thereof) \rev{according to the} question. For example, in question T01, "Could this be a disk viewed edge-on?", we see there are two possible answers for individuals to choose from: yes and no. Therefore the votes are stored in two categories, those of "yes" and those of "no". If 25\% of the population votes that a specific galaxy could be viewed edge-on, \revthree{an} answer of "yes" according to the question, then the voting fraction of "edge-on" would be 0.25, and the voting fraction of "not edge-on" is 0.75. All answers to a specific question, when added together, must have a voting fraction of 1, which represents 100\% of the population that answered the question.

As a result of the decision tree and tiered questions, not all participants will answer each question; higher tiers, denoted by color on Figure~\ref{fig:gz_tree}, may have fewer votes than the grey tiers that each participant answers. For example, the 4th tier questions will only be answered by participants that vote in favor of a galaxy having features, being face-on, and appearing to have a spiral pattern. This means that if we start with a small sample size, higher-tier questions run into the realm of small-number statistics. 

As a citizen science project, \rev{it is important to note that Galaxy Zoo can be susceptible to human bias. However,} with extensive available data, Galaxy Zoo has been used in conjunction with GAMA \rev{to minimize this bias and take full advantage of the data.} \rev{Such studies include identifying} dust lanes in edge-on galaxies \citep{Holwerda2019}, strong gravitational lensing \citep{Knabel2020}, green valley galaxy morphology \citep{Smith2022}, and \rev{investigating} a possible correlation between the number of spiral arms in spiral galaxies and star formation \citep{Porter-Temple2022}. 

\subsection{Sample Selection}
\label{sec:sampleselection}

\begin{figure*}
	\includegraphics[scale=0.7]{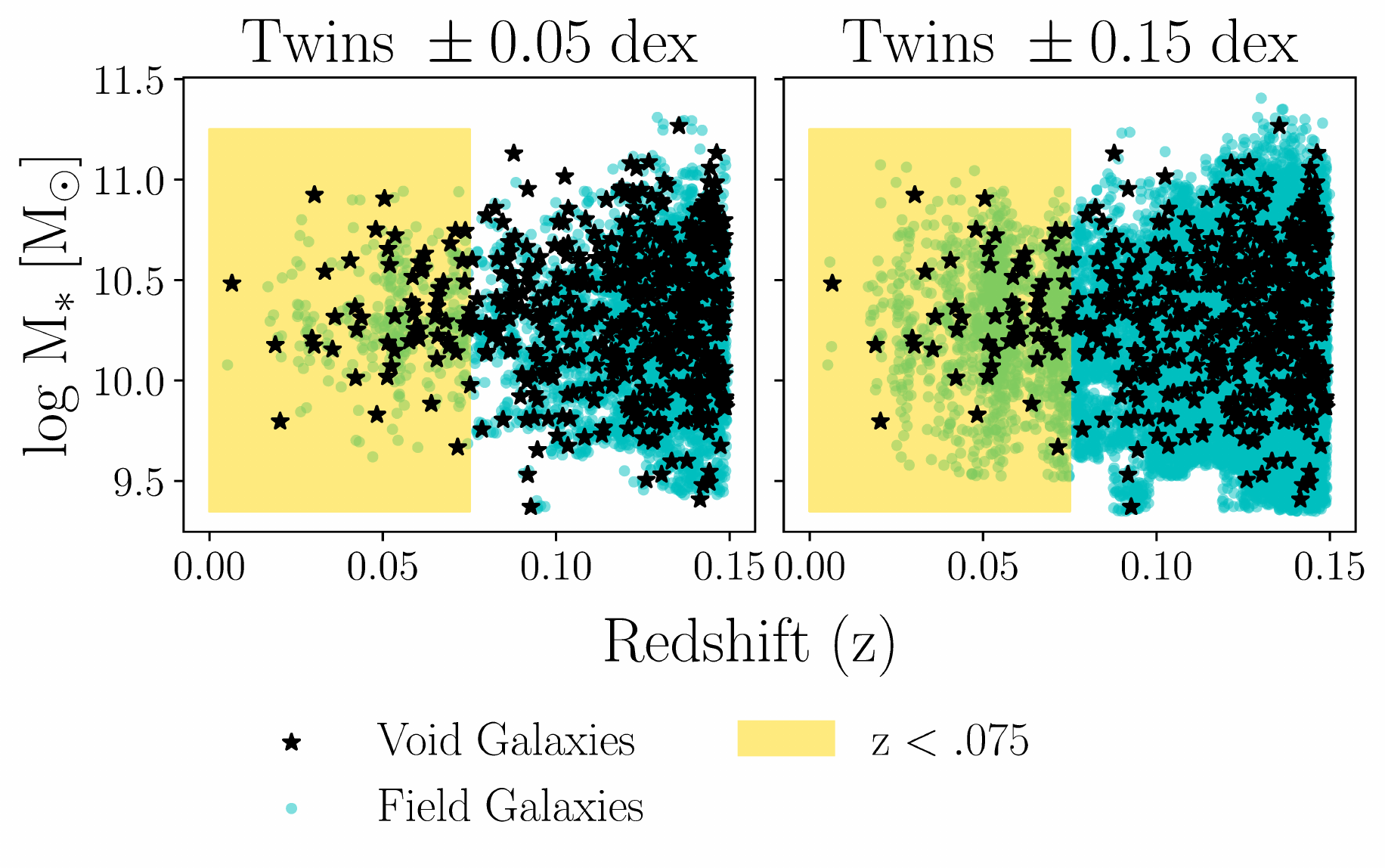}
    \caption{The complete samples of data from GAMA and GalaxyZoo. Void galaxies are denoted as black stars, whereas other galaxies in GAMA and GalaxyZoo are cyan circles. The yellow highlighted region represents the redshift-limited portion of this study, where left and right panels show the difference in the population of comparable field galaxies to void galaxies.  }
    \label{fig:completenessplot}
\end{figure*}

\begin{table*} 
\caption{\rev{Summary of the number of galaxies present for each sample, before conducting analysis. This table does not define the number of galaxies that exist in a certain morphology, for example, but instead the number of galaxies present in the cumulative population histograms beginning with Figure~\ref{fig:sersic_hist}. }Note that later in this study for question T09, to avoid redundancy, we simply combine the answers for "merging", "tidal debris", and "both."} 
\label{table:numbers}
\centering
    \begin{tabular}{ccccccc} \toprule
        \toprule
        &  \multicolumn{3}{c}{\rev{0 < z < 0.075}} & \multicolumn{3}{c}{\rev{0.075 < z < 0.15}} \\
        \cmidrule(lr){2-4} \cmidrule(lr){5-7} 
          & Void Galaxies & Twins ($\pm0.05$ dex) & Twins ($\pm0.15$ dex) 
          & Void Galaxies & Twins ($\pm0.05$ dex) & Twins ($\pm0.15$ dex)  \\ 
        \midrule 
        \rev{S\'ersic} Index, Effective Radius 
        & 58 & 350 & 1334 & 444 & 4831 & 15710  \\
        T00: Features 
        & 57 & 349 & 1325 & 436 & 4614 & 14843 \\
        T02: Bar 
        & 43 & 240 & 860 & 227 & 1986 & 6116 \\
        T03: Spiral 
        & 36 & 199 & 758 & 242 & 2230 & 6618 \\
        T04: No Central Bulge 
        & 26 & 179 & 652 & 178 & 2721 & 8772 \\
        T04: Obvious Central Bulge 
        & 50 & 320 & 1224 & 308 & 3896 & 12486 \\
        T04: Dominant Central Bulge 
        & 50 & 288 & 1133 & 289 & 3518 & 11444  \\
        T07: Edge-on: Rounded Bulge 
        & 29 & 338 & 1252 & 159 & 4293 & 13918 \\
        T07: Edge-on: Boxy Bulge 
        & 15 & 147 & 603 & 110 & 2728 & 9077 \\
        T07: Edge-on: No Bulge 
        & 19 & 209 & 826 & 112 & 3096 & 10128 \\
        T09: Evidence of Mergers 
        & 55 & 350 & 1334 & 425 & 4828 & 15702 \\
        \bottomrule
    \end{tabular}
\end{table*}

To conduct our analysis, it is first important to ensure that we are only investigating a range of galaxies in which we are sure that our samples of both field and void galaxies are complete. Because Galaxy Zoo has a maximum redshift of $\rm z_{max}$ = 0.15, our \rev{maximum} redshift of this sample is \revtwo{also} limited to $\rm z_{max} = 0.15$. Furthermore, we limit our total stellar mass range to that of $\rm 10^{9.25} M_{\odot} < M_* < 10^{11.25} M_{\odot}$ \rev{, as this mass range is home to our identified void galaxies, and is most easily compared to previous literature on void galaxies.}

\rev{This} study's primary focus is to determine whether void galaxies and similar field galaxies have a differing average morphology. This means that we are attempting to test for significance in the two samples where the primary difference is the environment (void vs field).

\rev{It is known that morphological features in galaxies can be redshift-dependent; galaxies residing around z=0 are a different population than those at z=0.1.} To ensure we are taking \rev{redshift} into consideration while maintaining an appropriate sample size to allow for reasonable statistics, we divide our sample into two: one consisting of \rev{ 0 < z < 0.075} (yellow shaded region of Figure~\ref{fig:completenessplot}) and another \rev{of 0.075 < z < 0.15 (unshaded portion of Figure~\ref{fig:completenessplot}). We can now effectively refer to these samples as our "local" galaxies (0 < z < 0.075) and "distant" galaxies (0.075 < z < 0.15), similar to \citet{Rojas2004}. }

The \rev{0 < z < 0.075 sample will be important when analyzing voting fractions of morphologies such as the presence of a bar (questions T02 and T07) or tidal debris (question T09), as \citet{Kruk2018} \revtwo{find} that few bars are accurately resolved above a redshift of z = 0.1, and therefore limit their sample for bars to z = 0.06. Similarly, \citet{Porter-Temple2022}, which utilizes the same GAMA and Galaxy Zoo data to investigate the number of spiral arms (another morphological feature), \rev{limit} their sample to $\rm z_{max}$ = 0.08. These redshift cuts ensure that the data gathered by Galaxy Zoo is from sufficiently resolved galaxy images. Our study is slightly more complicated in the fact that we investigate a wide range of morphological components, some of which do not require such precise resolution, such as features (question T00), spiral arm patterns (question T03), and discerning between the presence of a bulge or not (questions T04 and T07). } 

\rev{However, because we are interested in ensuring that the primary difference between our void and field galaxies is their environment, we further limit the sample of field galaxies by identifying directly comparable galaxies, which we refer to as \revthree{`}twins' of the void galaxies. In terms of redshift, we acknowledge that the \revthree{`}distant' 0.075 < z < 0.15 is still a large redshift range, and therefore require that, in order for a field galaxy to be identified as a twin to a void galaxy, it must have a redshift within $\pm0.025$ of an identified void galaxy. }

\rev{We define star formation rates and stellar mass to be equally important in identifying void galaxy analogues. Therefore, \revthree{`}twins' are also required to be any field galaxy that is within $\pm 0.05$ dex or $\pm 0.15$ dex of a void galaxy in terms of sSFR and $\rm M_*$.} The intention behind this is to identify a small subset of galaxies that are almost exactly identical to the void galaxies ($\pm 0.05$ dex) in terms of properties, but due to observational uncertainties in terms of properties such as sSFR, we allow for the second, larger sample of comparable field galaxies ($\pm 0.15$ dex). Keeping both definitions of 'twins' is important to ensure we are maintaining similar samples, all while providing an appropriate number of field galaxies to compare with the void galaxies (see Table \ref{table:numbers}). 

\rev{In summary, we have two samples of void galaxies and their field galaxy \revthree{`}twins': 0 < z < 0.075, and 0.075 < z < 0.15. The former redshift range requires that, to be a \revthree{`}twin', a field galaxy must have a sSFR and $\rm M_*$ within $\pm0.05$ or $\pm0.15$ dex of a void galaxy. The latter redshift sample implements the same sSFR and $\rm M_*$ requirement, but imposes the additional restraint that the field galaxy is \textit{also} within $\pm 0.025$ in redshift of the same galaxy. If a field galaxy does not meet all requirements for a specific void galaxy, it will not be identified as a \revthree{`}twin'.} \rev{To remain complete, we later conduct an analysis and statistical significance testing on all subgroups.}

\rev{The overall numbers of the galaxies within our analysis (void galaxies, field galaxies within $\pm 0.05$ dex, and field galaxies within $\pm 0.15$ dex) are documented in Table \ref{table:numbers}. Note that in the Galaxy Zoo questions, this table does not provide the number of galaxies with that specific morphological feature (e.g., the bar voting fraction row does not say how many galaxies have bars), but rather the total number of galaxies for which we have voting results. }

\section{Results}
\label{sec:results}

After constraining our sample, we analyze and compare the physical properties of the void galaxies\rev{, and compare them to the field galaxy analogues.} The properties included here will be directly relevant to morphology: \rev{S\'ersic} index, specific star formation rate (sSFR), and effective radius. Once we understand the distribution of these components, we can look at specific morphological voting in Galaxy Zoo.

\subsection{Physical Properties}
\label{sec:physicalproperties}

\begin{figure*}
	\includegraphics[scale=.6]{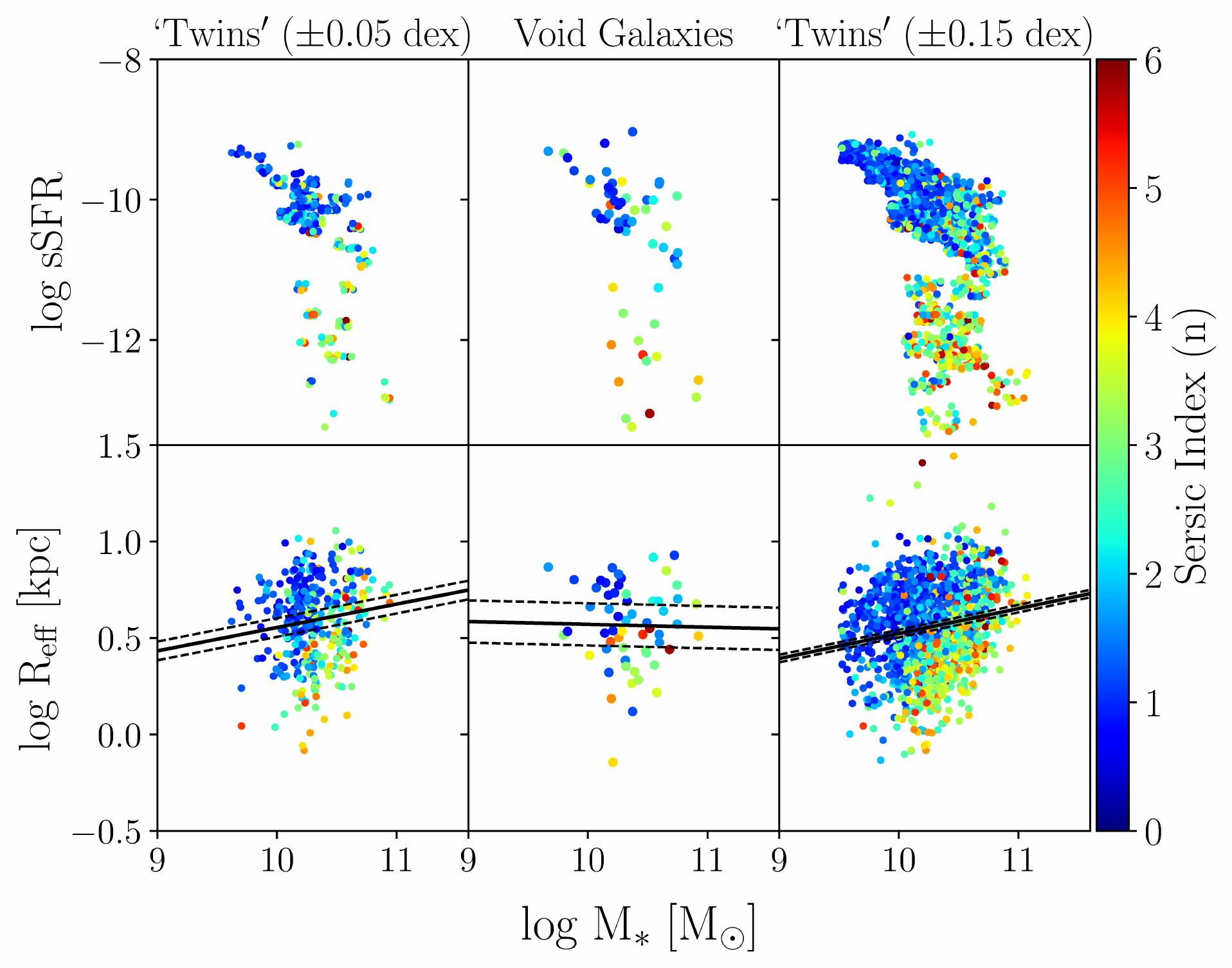}
    \caption{Physical properties of galaxies in both samples with redshift \rev{0 < z < 0.075,} GAMA \revthree{`}twins' within $\pm0.05$ dex (left panels) and $\pm0.15$ dex (right panels). The top panels show specific star formation rate (sSFR) as a function of stellar mass ($\rm M_{*}$), and the bottom panels show effective radius ($\rm R_{eff}$; kpc) as a function of $M_{*}$. Points are colored by their \rev{S\'ersic} index. \revthree{The solid black line in the bottom panels represents the least squares regression line from jacknife resampling, the equation and error for which can be found in Table~\ref{table:sizemass}. Dashed black lines represent the $\pm 1 \sigma$ error.}  }
    \label{fig:allprops_zcut}
\end{figure*}

The physical properties of the galaxy, including the rate at which they are actively producing stars and their effective radius, provide useful information about their history and distribution in the Universe. The following Figures, beginning with Figure~\ref{fig:allprops_zcut}, allow us to investigate these in our sample. 

\rev{ Figure~\ref{fig:allprops_zcut} displays the local sample (0 < z < 0.075).} In the top panels, it is clear that the diskier (n < 2; \rev{shades of blue} in Figure~\ref{fig:allprops_zcut}) galaxies have \rev{specific star formation rates that are nearly two orders of magnitude larger than the elliptical galaxies (n $\sim$ 4; \rev{yellow/green} in Figure~\ref{fig:allprops_zcut}) .} In the bottom panels, we see that these ellipticals have \revthree{similar effective radii to the disks, but maintain a similar or slightly higher (up to an order of magnitude) stellar mass. } \rev{Throughout all of Figure~\ref{fig:allprops_zcut}, but most evident in the top panels with sSFR, each \revthree{morphological} group appears to cluster together. While there is some slight variation, disk and elliptical galaxies are clearly separated in the 0 < z < 0.075 sample. }

\begin{figure*}
	\includegraphics[scale=.6]{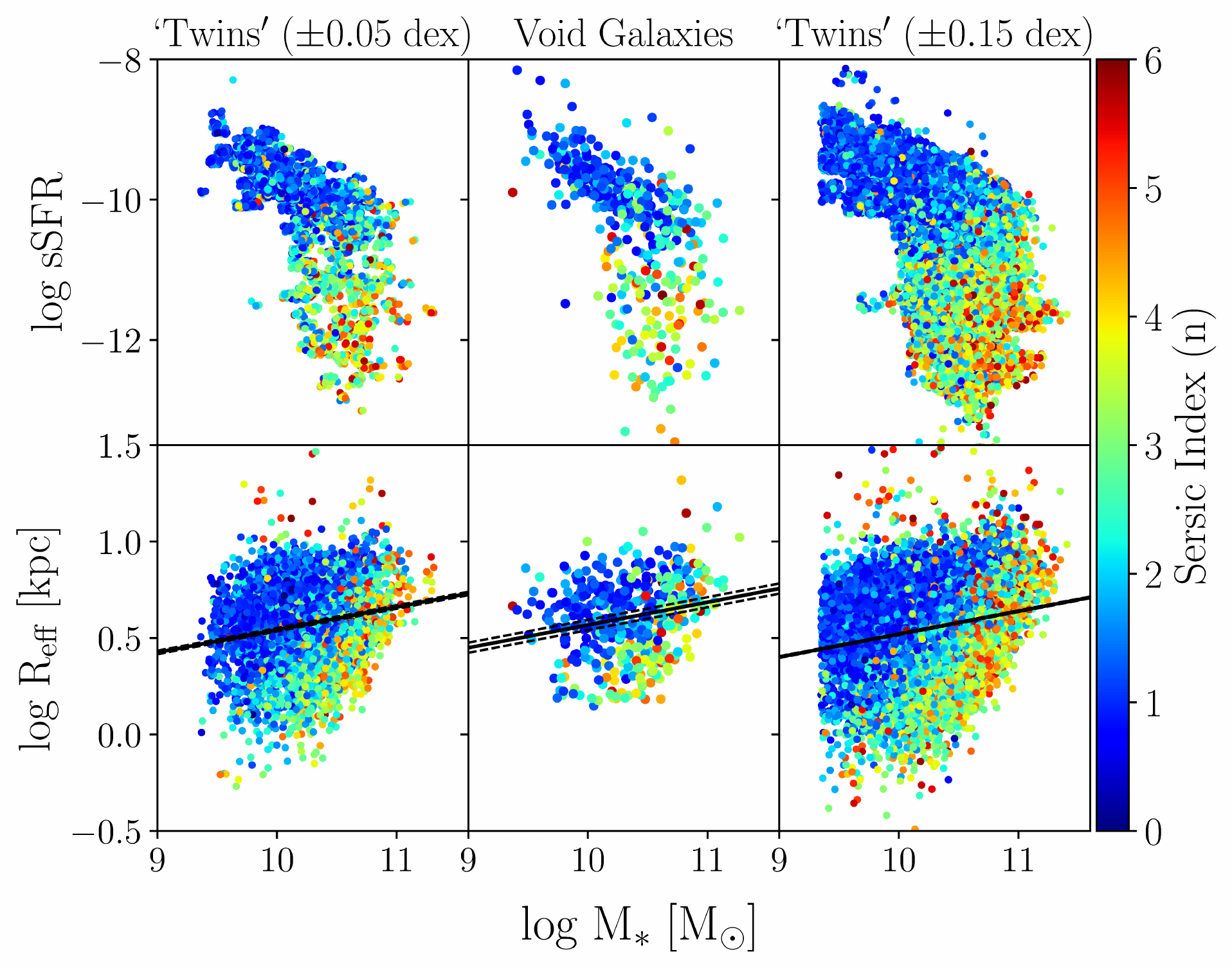}
    \caption{Physical properties of galaxies in both samples with redshift \rev{0.075 < z < 0.15}, GAMA \revthree{`}twins' within $\pm0.05$ dex (left panels) and $\pm0.15$ dex (right panels). The top panels show specific star formation rate (sSFR) as a function of stellar mass ($\rm M_{*}$), and the bottom panels show effective radius ($\rm R_{eff}$; kpc) as a function of $M_{*}$. Points are colored by their \rev{S\'ersic} index.  \revthree{The solid black line in the bottom panels represents the least squares regression line from jacknife resampling, the equation and error for which can be found in Table~\ref{table:sizemass}. Dashed black lines represent the $\pm 1 \sigma$ error.}}
    \label{fig:allprops_nozcut}
\end{figure*}

If we consider the sample of galaxies \revtwo{in the redshift range of} \rev{(0.075 < z < 0.15}; Figure~\ref{fig:allprops_nozcut}), we still see this result. While there is a significantly bigger population of galaxies due to the extended sample size, we can still clearly discern that in terms of sSFR (top panels), disk and elliptical galaxies reside in their own regimes. 

We note here that we include no analysis on the difference in sSFR between void and field galaxies, as we use this property to constrain our sample of field galaxies to those that are intrinsically similar to the sample of void galaxies. 

\subsubsection{Size-Mass Relation and Effective Radius}

\begin{table*} 
\caption{\revthree{Slope, error, y-intercept, and p-value for the least squares regression lines in the bottom panels of Figures~\ref{fig:allprops_zcut} and \ref{fig:allprops_nozcut}, obtained via jacknife resampling. Error is 1$\sigma$ in regards to the slope. Bold p-values represent significance for a relationship between stellar mass ($\rm M_*$) and effective radius ($\rm R_{eff}$)}. }
\label{table:sizemass}
\centering
    \begin{tabular}{ccccccc} \toprule
         \toprule & \multicolumn{3}{c}{\rev{0 < z < 0.075}} & \multicolumn{3}{c}{\rev{0.075 < z < 0.15}} \\

        \toprule & \multicolumn{1}{c}{Void Galaxies} & \multicolumn{1}{c}{`Twins' $\pm$0.05 dex} & \multicolumn{1}{c}{`Twins' $\pm$0.15 dex} &  \multicolumn{1}{c}{Void Galaxies} & \multicolumn{1}{c}{`Twins' $\pm$0.05 dex} & \multicolumn{1}{c}{`Twins' $\pm$0.15 dex} \\

        \midrule 
        \revthree{Slope}
        & 0.12 & -0.01 & 0.13 & 0.12 & 0.12 & 0.12  \\
        \revthree{Error}
        & $\pm$0.05 & $\pm$0.11 & $\pm$0.02 & $\pm$0.03 & $\pm$0.01 & 0.004 \\
        \revthree{Y-intercept}
        & -0.66 & 0.72 & -0.77 & -0.61 & -0.63 & -0.67  \\
        \revthree{P-value}
        & \textbf{0.01} & 0.88 & \textbf{<0.01} & \textbf{<0.01} & \textbf{<0.01} & \textbf{<0.01} \\
        \bottomrule
    \end{tabular}
\end{table*}

The size of the galaxies in question is a basic morphological property that can be telling about the galaxy's history.  As a result, the relationship between effective radius and stellar mass, often known as the galaxy size-mass relation, is thought to be another indicator of evolution in a galaxy \citep{VanDerWel2014, Genel2018, Mowla2019, Kawinwanichakij2021, Suess2021, Yang2021,  Nedkova2022}. Typically, this relation can be understood as larger galaxies tend to also be more massive, which is commonly thought to be a result of mergers \citep{Hernquist1989, Robertson2006, Naab2010}. In the bottom panels of Figures~\ref{fig:allprops_zcut} and \ref{fig:allprops_nozcut}, we use jacknife sampling to accurately fit the size-mass relation, represented by the solid black line.  

\begin{figure}
	\includegraphics[width=\columnwidth]{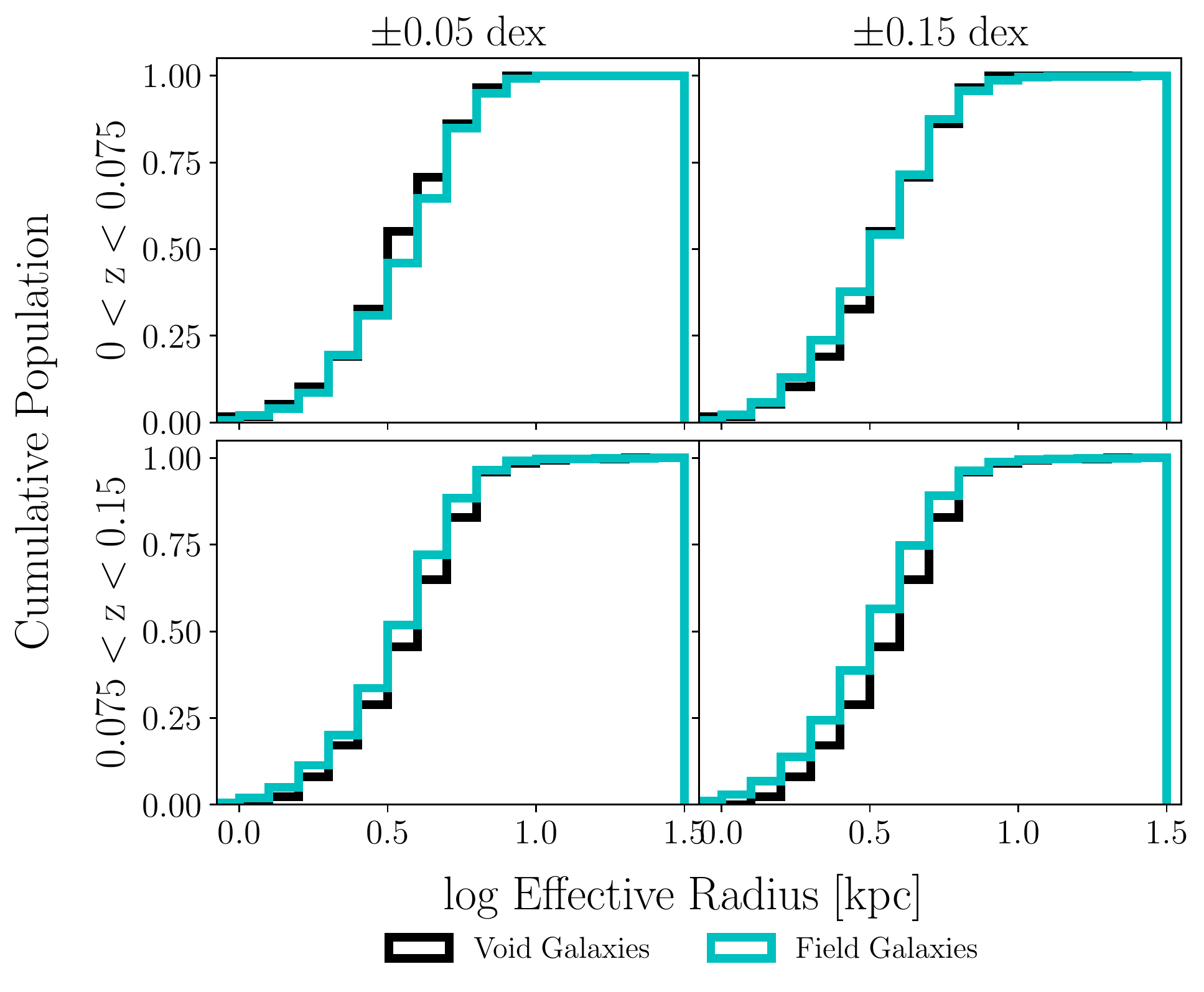}
    \caption{Cumulative histogram of effective radius values. Void galaxies are denoted in black, whereas twins of the void galaxies in GAMA are in cyan. Left panels represent twins chosen within $\pm0.05$ dex of void galaxies, whereas right panels represent twins chosen within $\pm0.15$ dex. \revthree{Most panels show that roughly half of their populations lie within $\rm R_{eff}$ of $10^{0.5}$ (3.16) kpc.}}
    \label{fig:effectiveradius_hist}
\end{figure}

Focusing specifically on the size of these galaxies, Figure~\ref{fig:effectiveradius_hist} \rev{represents} a similar cumulative histogram of the effective radii to see whether there is a difference in size between void galaxies and their field counterparts. \rev{Here, we see that most galaxies in both samples reside within 1-10 kpc, as expected. \revthree{It is interesting to note that, in Figure~\ref{fig:allprops_nozcut}, the line of best fit is nearly identical across the void galaxies and all samples of field galaxies, unlike Figure~\ref{fig:allprops_zcut}, though we note the importance of sample size.} When conducting Kolmogorov-Smirnov (K-S) testing, we \revthree{only find significance in the $\pm0.05$ dex, 0.075 < z < 0.15, sample (see Table~\ref{table:kstest}, Figure~\ref{fig:effectiveradius_hist})}.}  

\subsubsection{\rev{S\'ersic} Index}

The \rev{S\'ersic} index is one of the simplest ways to gain insight into the morphological distribution of galaxies. Plotting histograms of these values, \revtwo{calculated by \citet{Kelvin2012},} allows us to immediately see what the general distribution of galaxy morphology based on the light profile appears to be, with disky galaxies residing around n < 2, and ellipticals around n $\sim$ 4. In addition, this allows for a direct and normalized analysis between the samples of void and field galaxies. 

\begin{figure}
	\includegraphics[width=\columnwidth]{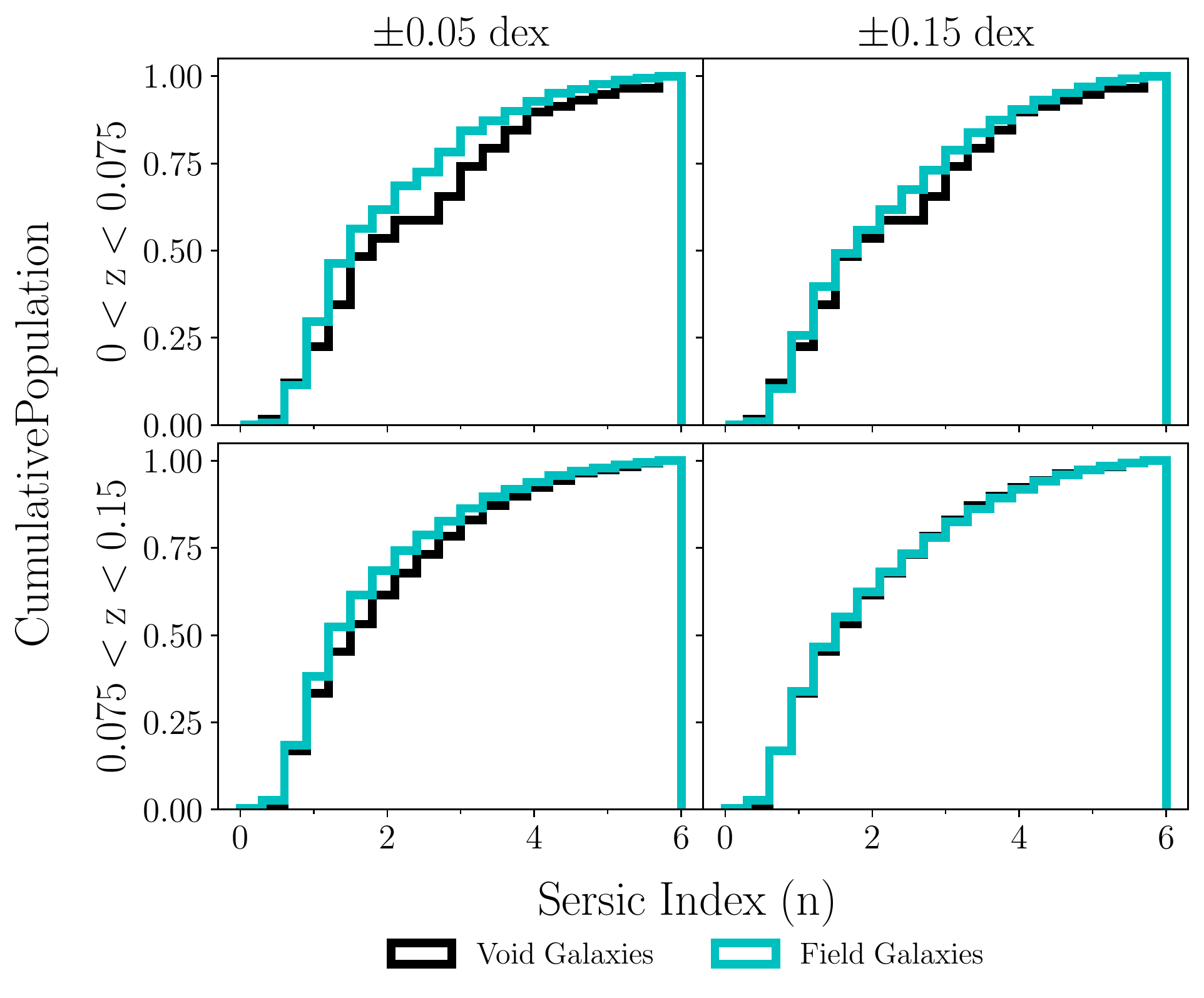}
    \caption{Histogram of \rev{S\'ersic} index (n) values. Void galaxies are denoted in black, whereas twins of the void galaxies in GAMA are in cyan. Left panels represent twins chosen within $\pm0.05$ dex of void galaxies, whereas right panels represent twins chosen within $\pm0.15$ dex. \revthree{All panels show that about 50\% of the their populations have a S\'ersic index of n < 2, and 75\% with n < 3, showing disk-dominated samples, as ellipticals are n $\approx$ 4. }}
    \label{fig:sersic_hist}
\end{figure}

For each population of galaxies in Figure~\ref{fig:sersic_hist}, we see a clearly defined peak in the distributions of \rev{S\'ersic} index at n < 2, with all subsamples having roughly \rev{half} of their galaxies with a \rev{S\'ersic} index of n < 2\rev{, and 75\% with n < 3,} showing that most galaxies in each distribution appear to be late-type, or disky. Therefore, we immediately see that both void galaxies and their \revthree{`}twins' in redshift, stellar mass, and sSFR are disk-dominated. This fact is not changed whether we look at the $\pm$0.05 dex (left panels of Figure~\ref{fig:sersic_hist}) or $\pm$0.15 dex samples of twins (right panels). \revthree{While for the \revthree{`}local' sample we need to be careful in over-interpreting results due to the smaller sample size of void galaxies, we do find the differences to be statistically significant for both subsamples of field galaxies within 0.075 < z < 0.15 (upper panels of Figure~\ref{fig:sersic_hist}; see also Table~\ref{table:kstest}). }

\subsection{GalaxyZoo}

\begin{table*} 
\caption{Significance testing results \revtwo{using the two-sample Kolmogorov-Smirnov test} for morphological features between void galaxies and their twins, under the null hypothesis that both samples come from the same distribution. Bold p-values denote significant results \revtwo{(p-value < 0.05)}. }
\label{table:kstest}
\centering
    \begin{tabular}{ccccccccc} \toprule
         \toprule & \multicolumn{4}{c}{\rev{0 < z < 0.075}} & \multicolumn{4}{c}{\rev{0.075 < z < 0.15}} \\

        \toprule &  \multicolumn{2}{c}{$\pm$0.05 dex} & \multicolumn{2}{c}{$\pm$0.15 dex} &  \multicolumn{2}{c}{$\pm$0.05 dex} & \multicolumn{2}{c}{$\pm$0.15 dex} \\

        \cmidrule(lr){2-3} \cmidrule(lr){4-5} \cmidrule(lr){6-7} \cmidrule(lr){8-9}  
          & Test Statistic & P-value & Test Statistic & P-value 
          & Test Statistic & P-value & Test Statistic & P-value\\ 
        \midrule 
        \rev{S\'ersic} Index 
        & 0.10 & 0.66 & 0.07 & 0.94 & 0.09 & \textbf{<0.01} & 0.13 & \textbf{<0.01} \\
        Effective Radius 
        & 0.14 & 0.24 & 0.09 & 0.72 & 0.09 & \textbf{<0.01} & 0.03 & 0.88 \\
        T00: Features 
        & 0.06 & 0.99 & 0.11 & 0.45 & 0.10 & \textbf{<0.01} & 0.13 & \textbf{<0.01} \\
        T02: Bar 
        & 0.09 & 0.93 & 0.08 & 0.92 & 0.08 & 0.14 & 0.10 & \textbf{0.02} \\
        T03: Spiral 
        & 0.25 & \textbf{0.04} & 0.27 & \textbf{0.01} & 0.08 & 0.09 & 0.10 & \textbf{0.01} \\
        T04: No Central Bulge 
        & 0.27 & \textbf{0.05} & 0.33 & \textbf{<0.01} & 0.39 & \textbf{<0.01} & 0.42 & \textbf{<0.01} \\
        T04: Obvious Central Bulge 
        & 0.25 & \textbf{0.01} & 0.18 & 0.08 & 0.28 & \textbf{<0.01} & 0.30 & \textbf{<0.01} \\
        T04: Dominant Central Bulge 
        & 0.11 & 0.60 & 0.13 & 0.33 & 0.29 & \textbf{<0.01} & 0.32 & \textbf{<0.01} \\
        T07: Edge-on: Rounded Bulge 
        & 0.43 & \textbf{<0.01} & 0.46 & \textbf{<0.01} & 0.60 & \textbf{<0.01} & 0.62 & \textbf{<0.01} \\
        T07: Edge-on: Boxy Bulge 
        & 0.37 & \textbf{0.03} & 0.42 & \textbf{0.01} & 0.57 & \textbf{<0.01} & 0.60 & \textbf{<0.01} \\
        T07: Edge-on: No Bulge 
        & 0.37 & \textbf{0.01} & 0.40 & \textbf{<0.01} & 0.56 & \textbf{<0.01} & 0.59 & \textbf{<0.01}  \\
        T09: Evidence of Mergers 
        & 0.13 & 0.37 & 0.13 & 0.27 & 0.09 & \textbf{<0.01} & 0.11 & \textbf{<0.01}  \\
        \bottomrule
    \end{tabular}
\end{table*}

\begin{figure}
	\includegraphics[width=\columnwidth]{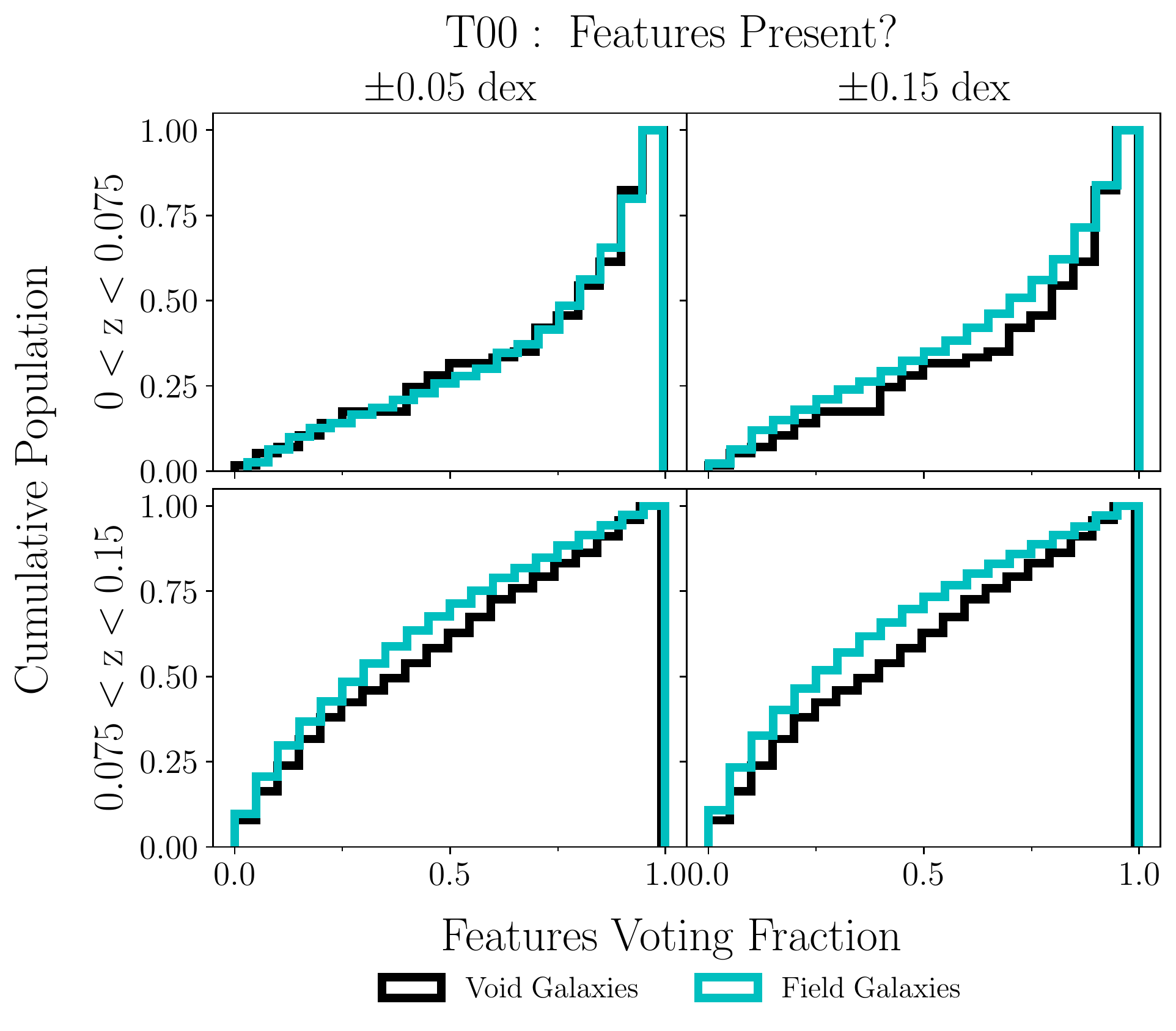}
    \caption{Histogram comparing the voting fraction for question T00 (presence of features) in both samples, with a normalized frequency. GAMA \revthree{`}twins' are denoted in cyan, and void galaxies are denoted in black. Left panels represent twins chosen within $\pm0.05$ dex of void galaxies, whereas right panels represent twins chosen within $\pm0.15$ dex. \revtwo{Note that values of \revthree{`}0' mean that no \revthree{citizen scientists} answered the question with \revthree{`}yes', while \revthree{`}1' means all answered \revthree{`}yes', or in favor of the specific morphological component (i.e., features).} \revthree{In the local sample (top panels), 75\% of the population has voting fractions greater than 0.5, indicating that a majority of the galaxies have features present, while the further sample (bottom panels) only have 25-50\% of their population in the same range, meaning that features are much less common in this further redshift range.} }
    \label{fig:features_hist}
\end{figure}

Similar to the \rev{S\'ersic} index and effective radius, we now investigate the voting fractions from the selected Galaxy Zoo questions. Here, we focus our attention on the following questions from Figure~\ref{fig:gz_tree}:

 \begin{itemize}
     \item T00 - Is the galaxy in the centre of the image simply smooth and rounded, or does it have features?
     \item T02: Is there any sign of a bar feature through the centre of the galaxy?
     \item T03: Is there any sign of a spiral arm pattern?
     \item T04: How prominent is the central bulge, compared with the rest of the galaxy?
     \item T07: Does the galaxy have a bulge at its centre?
     \item T09: Is the galaxy currently merging or is there any sign of tidal debris?
  \end{itemize}

We choose to skip question T01 (\rev{"}Could this be a disk viewed edge-on?\rev{"}) because edge-on galaxies are not a type of morphology that can be caused by environment; edge-on galaxies are merely a result of the viewing angle, so this specific question is not relevant to this study. Therefore, we skip to question T07 which contains the morphological information for galaxies viewed at such an angle. 
 
We also choose to skip questions T05 (\rev{"}How tightly wound do the spiral arms appear?\rev{"}) and T06 (\revtwo{"}How many spiral arms are there?\rev{"}) because we are simply interested in whether the spiral morphology itself is present as opposed to the intricacies involved. 

\begin{figure}
	\includegraphics[width=\columnwidth]{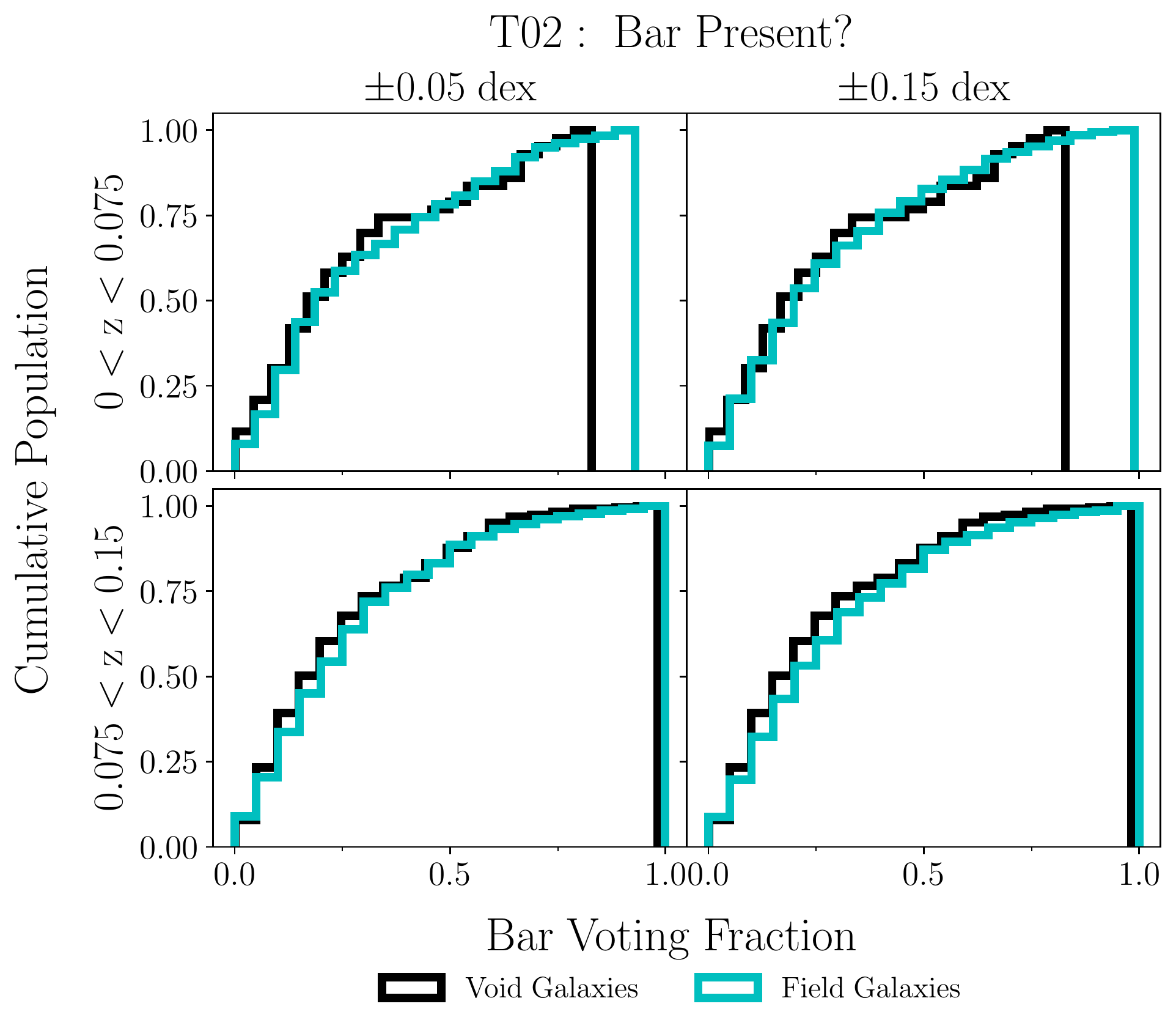}
    \caption{Histogram comparing the voting fraction for question T02 (\rev{"}Is a bar present?\rev{"}) in both samples, with a normalized frequency. GAMA \revthree{`}twins' are denoted in cyan, and void galaxies are denoted in black. Left panels represent twins chosen within $\pm0.05$ dex of void galaxies, whereas right panels represent twins chosen within $\pm0.15$ dex. \revthree{In all samples, 75\% of the population has voting fractions less than 0.5, so most galaxies here do not have a visible bar.} }
    \label{fig:bar_hist}
\end{figure}

Figure~\ref{fig:features_hist} is the beginning of our comparisons in Galaxy Zoo with question T00 (\rev{"}Is the galaxy in the centre of the image simply smooth and rounded, or does it have features?\rev{"}). Immediately, we can see that the samples are \revtwo{relatively} similar. \revthree{In the near sample, we can clearly tell that both the field and void galaxies within our physical parameters are dominated by the presence of features, especially compared to the galaxies at higher redshifts (0.075 < z < 0.15). When we conduct the two-sample Kolmogorov-Smirnov (K-S) test between the void and field galaxies (see Table~\ref{table:kstest}), we see significant results in this further redshift range, but low test statistics, indicating that we can be confident in the samples' similarity. }

\begin{figure}
	\includegraphics[width=\columnwidth]{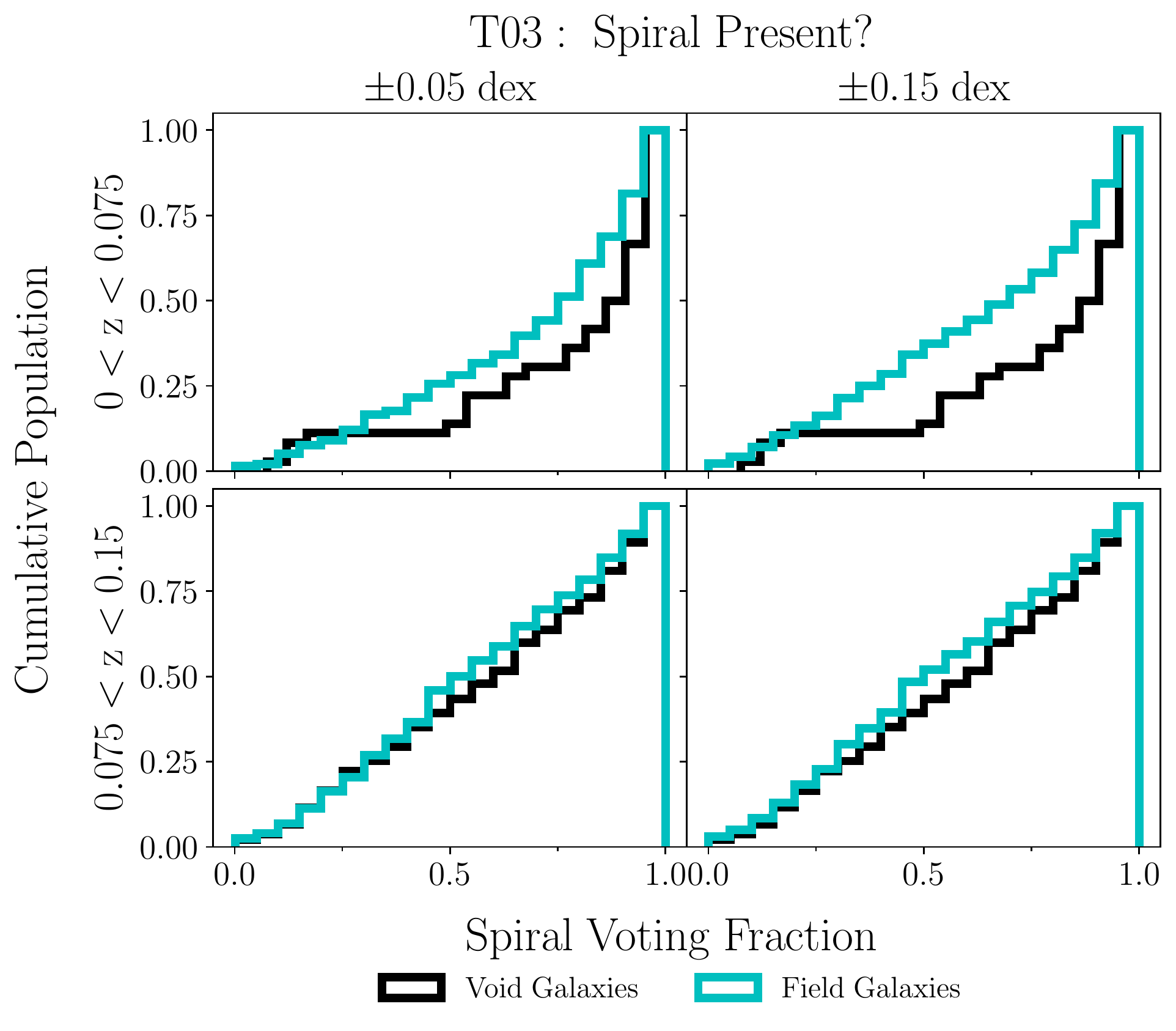}
    \caption{Histogram comparing the voting fraction for question T03 (presence of spiral \revtwo{arm pattern}) in both samples, with a normalized frequency. GAMA \revthree{`}twins' are denoted in cyan, and void galaxies are denoted in black. Left panels represent twins chosen within $\pm0.05$ dex of void galaxies, whereas right panels represent twins chosen within $\pm0.15$ dex. \revthree{In the top panels, about 75\% of each population has voting fractions above 0.5, showing majority spirals. The bottom panels are more evenly distributed, with roughly half of the populations at a voting fraction above 0.5.} }
    \label{fig:spiral_hist}
\end{figure}

Figures~\ref{fig:bar_hist} and \ref{fig:spiral_hist} address questions T02 and T03, which ask about the presence of a bar or spiral, respectively. Here we again note the importance of consulting the local region (0 < z < 0.075; top panels of Figure~\ref{fig:bar_hist}) for the presence of bars, as bars are not well resolved at higher redshifts, and \citet{Kruk2018} similarly limited their sample to z=0.06. \revthree{In the case of bars, we see low test statistics across all subsamples, and find significance in the $\pm0.15$ dex, 0.075 < z < 0.15 subsample. This indicates almost no difference in the presence of bars in void galaxies and field galaxies. This is an interesting result in itself, as bars may be formed by secular processes, yet we find no difference between the two galaxy populations. In Figure~\ref{fig:spiral_hist}, we see that spirals dominate both the void and field galaxie. However, at redshifts of 0 < z < 0.075 (upper panels), void galaxies seem to have a higher fraction of spirals. This is supported by the K-S test, which reveals moderate test statistics ($\sim$0.25) for the local group, and low test statistics for the further group ($\sim$0.09), including significance for three of the four subsamples.}

\begin{figure}
	\includegraphics[width=\columnwidth]{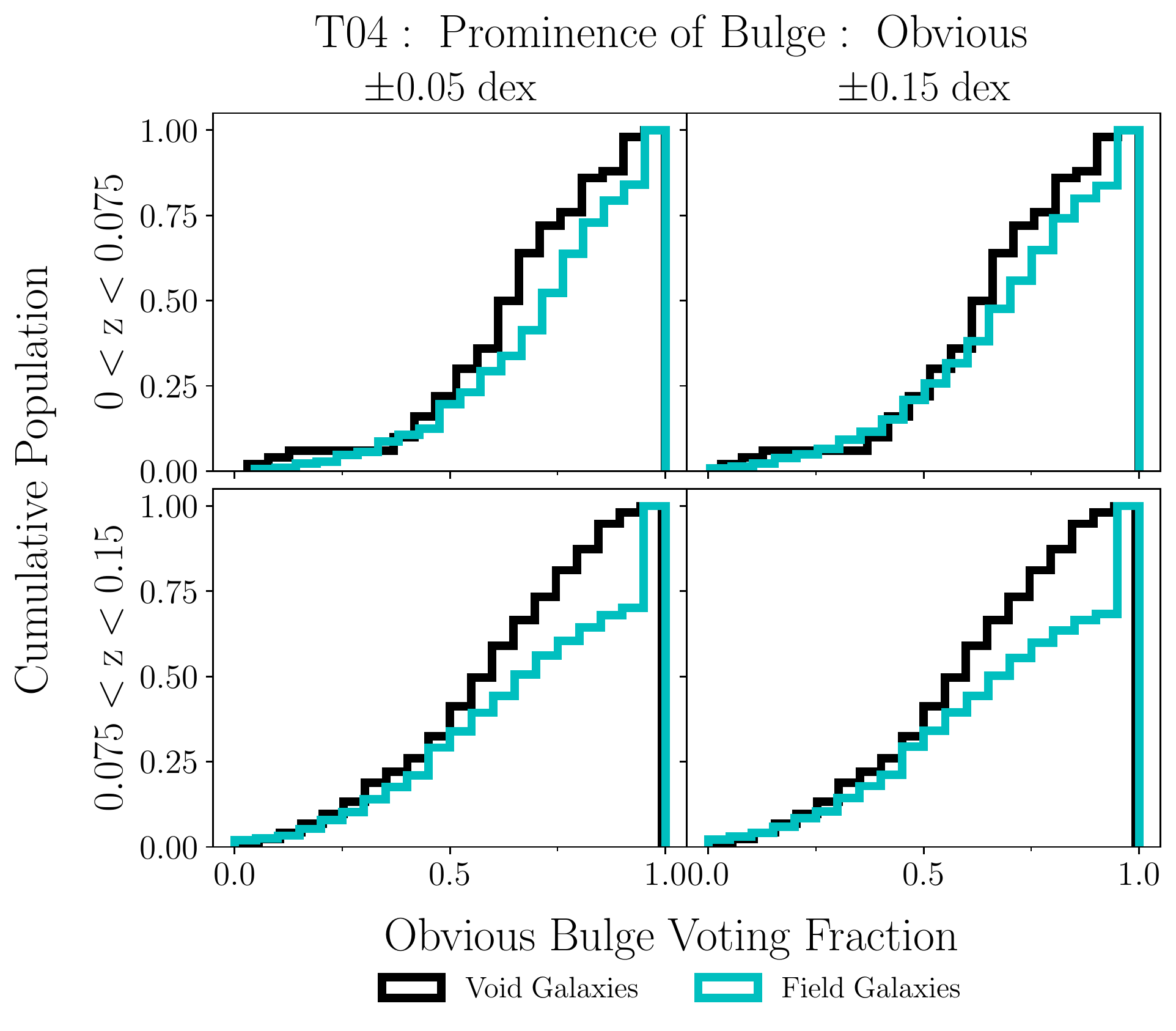}
    \caption{Histogram comparing the voting fraction for question T04 (\rev{"}How prominent is the central bulge, compared with the rest of the galaxy?\rev{"}) with answers for \revthree{`}obvious bulge' in both samples, with a normalized frequency. GAMA \revthree{`}twins' are denoted in cyan, and void galaxies are denoted in black. Left panels represent twins chosen within $\pm0.05$ dex of void galaxies, whereas right panels represent twins chosen within $\pm0.15$ dex. \revthree{Each sample has 75+\% of their voting fractions above 0.5, so most bulges can be classified as obvious.} }
    \label{fig:obviousbulge_hist}
\end{figure}

\begin{figure}
	\includegraphics[width=\columnwidth]{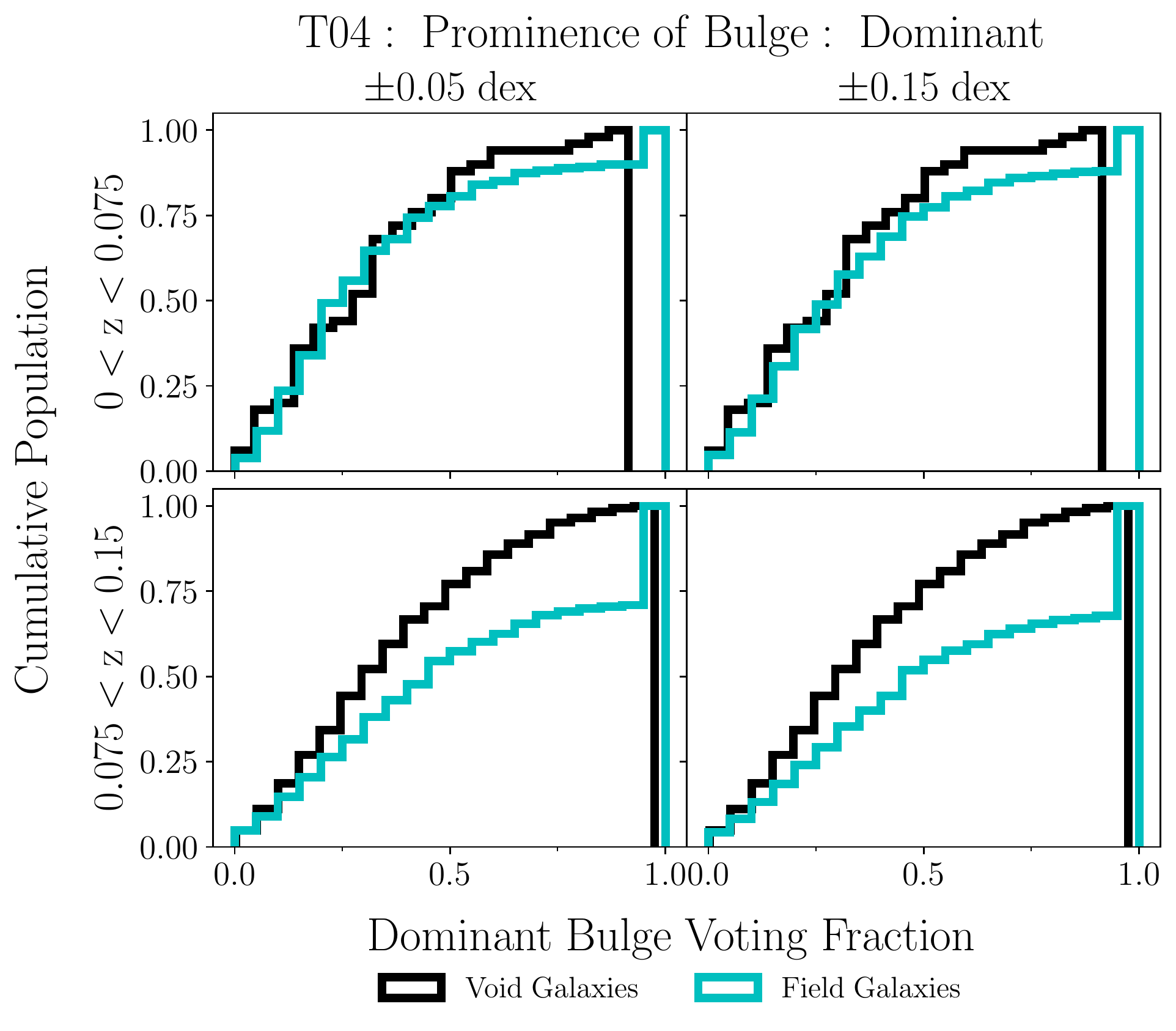}
    \caption{Histogram comparing the voting fraction for question T04 (\rev{"}How prominent is the central bulge, compared with the rest of the galaxy?\rev{"}) with answers for \revthree{`}dominant bulge' in both samples, with a normalized frequency. GAMA \revthree{`}twins' are denoted in cyan, and void galaxies are denoted in black. Left panels represent twins chosen within $\pm0.05$ dex of void galaxies, whereas right panels represent twins chosen within $\pm0.15$ dex. \revthree{Galaxies in the local regime (top panels) do not appear to have dominant bulges, nor do field galaxies in the further redshift range (bottom panels). Field galaxies in the latter regime appear to be evenly split.} }
    \label{fig:dominantbulge_hist}
\end{figure}

\begin{figure}
	\includegraphics[width=\columnwidth]{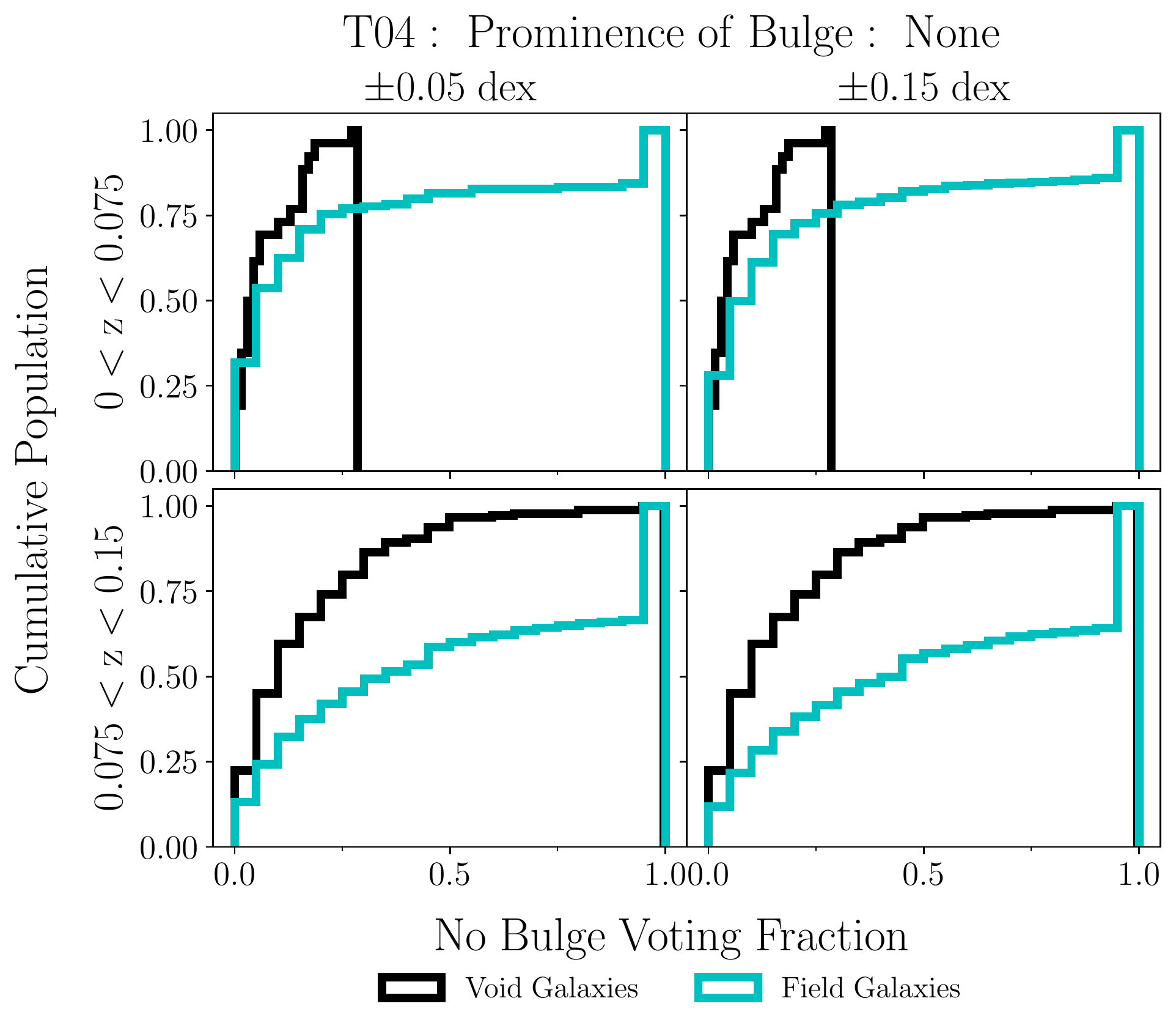}
    \caption{Histogram comparing the voting fraction for question T04 (\rev{"}How prominent is the central bulge, compared with the rest of the galaxy?\rev{"}) with answers for \revthree{`}no bulge' in both samples, with a normalized frequency. GAMA \revthree{`}twins' are denoted in cyan, and void galaxies are denoted in black. Left panels represent twins chosen within $\pm0.05$ dex of void galaxies, whereas right panels represent twins chosen within $\pm0.15$ dex. \revthree{All samples in the top panels have 75\% of their population within a voting fraction of 0.25, meaning participants strongly disagree with there being no bulge. In the bottom panels, this remains true for void galaxies, but field galaxies appear to be evenly split.} }
    \label{fig:nobulge_hist}
\end{figure}

Figures~\ref{fig:obviousbulge_hist}, \ref{fig:dominantbulge_hist}, and \ref{fig:nobulge_hist} represent the answers for question T04, which asks about the prominence of the central galaxy bulge compared to the rest of the galaxy (for those not identified as edge-on). \rev{Test statistics for this question are higher, suggesting the first difference in void and field galaxies is the prominence of the bulge. \revthree{In particular, we can note that the consensus of Figure~\ref{fig:nobulge_hist} is that void galaxies in all samples have a bulge present. All three of these questions appear to be highly significant at 0.075 < z < 0.15, but lose some of their significance in the local sample. This could be due to a variety of factors, including the limited sample size for lower redshifts. Question T04 likely needs higher-resolution images to determine an accurate answer.} A larger sample size and highly-resolved images would be best to follow-up on the dominance of central bulges in field and void galaxies, particularly to determine whether this is, in fact, a resolution issue, or whether this is a fundamental morphological difference between void and field galaxies at higher redshifts.}

\begin{figure}
	\includegraphics[width=\columnwidth]{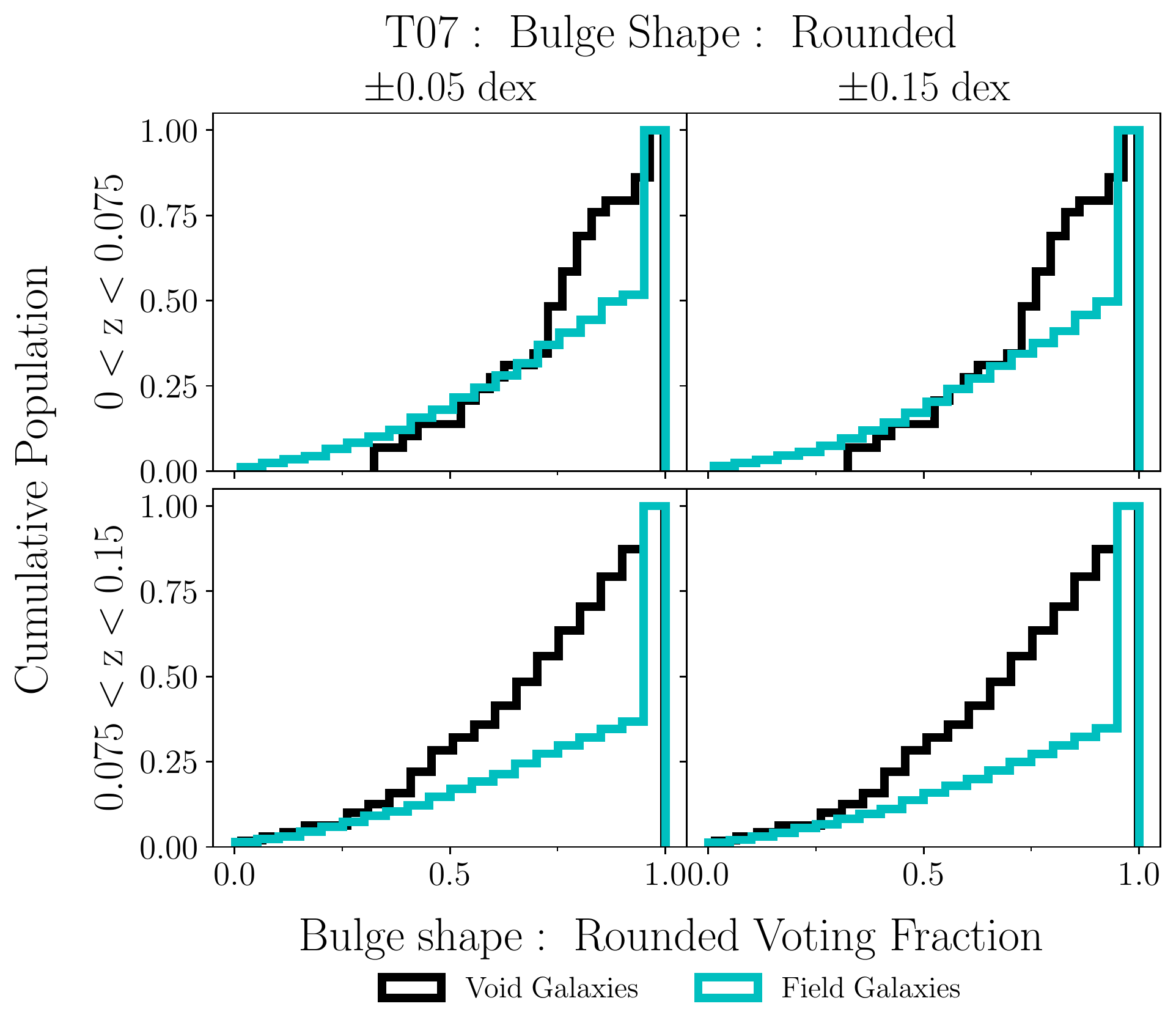}
    \caption{Histogram comparing the voting fraction for question T07 about edge-on galaxies (\rev{"}Does the galaxy have a bulge at its center?\rev{"}) with answers for \revthree{`}rounded bulge' in both samples, with a normalized frequency. GAMA \revthree{`}twins' are denoted in cyan, and void galaxies are denoted in black. Left panels represent twins chosen within $\pm0.05$ dex of void galaxies, whereas right panels represent twins chosen within $\pm0.15$ dex. \revthree{75+\% of all samples have a voting fraction greater than 0.5, indicating participants largely agree with the edge-on bulge being rounded.} }
    \label{fig:bulgeshaperounded_hist}
\end{figure}

\begin{figure}
	\includegraphics[width=\columnwidth]{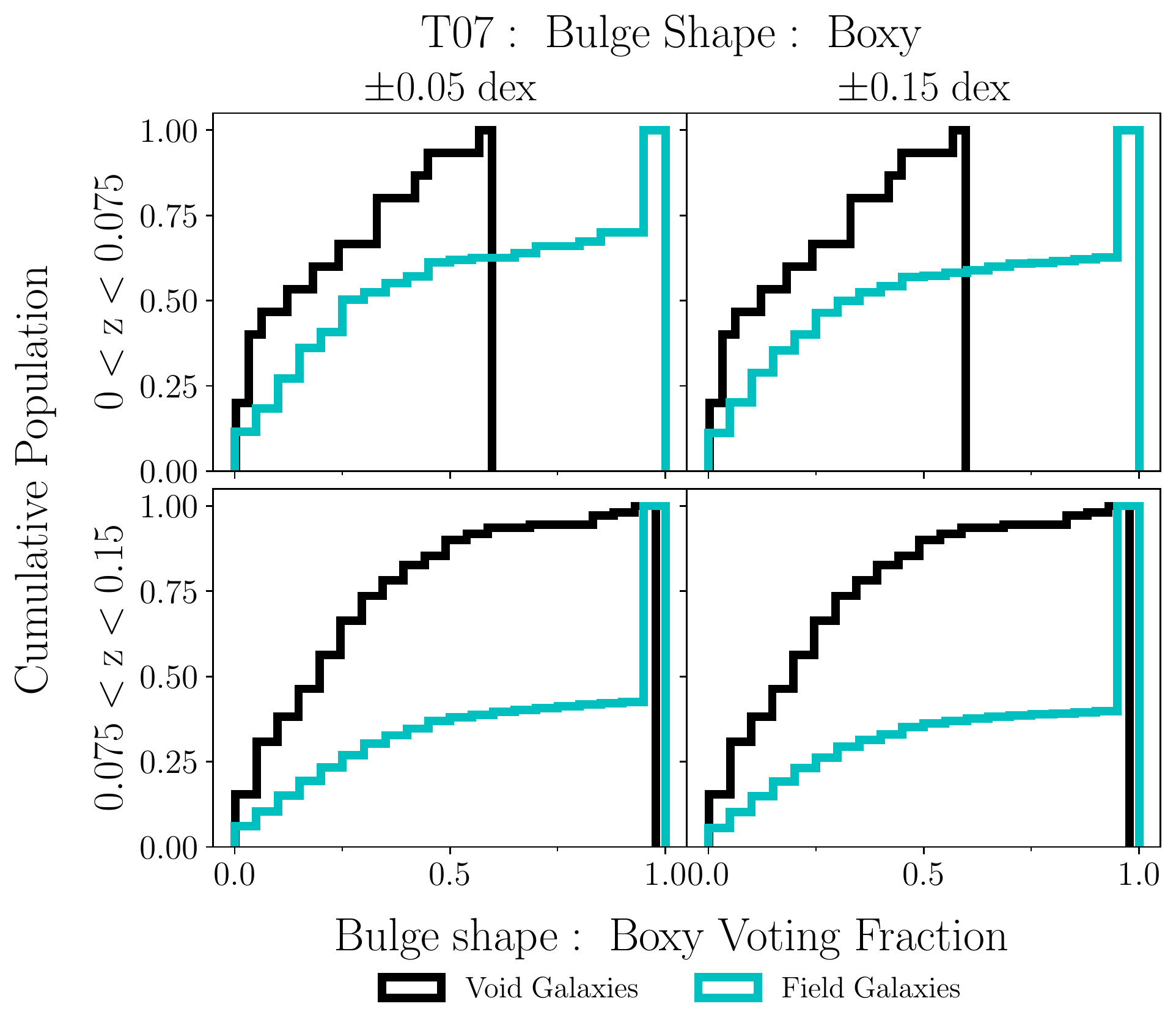}
    \caption{ Histogram comparing the voting fraction for question T07 about edge-on galaxies (\rev{"}Does the galaxy have a bulge at its center?\rev{"}) with answers for \revthree{`}boxy bulge' in both samples, with a normalized frequency. GAMA \revthree{`}twins' are denoted in cyan, and void galaxies are denoted in black. Left panels represent twins chosen within $\pm0.05$ dex of void galaxies, whereas right panels represent twins chosen within $\pm0.15$ dex. \revthree{In all panels, most void galaxies (75-90\%) do not have voting fractions that represent the presence of a boxy bulge. More than 25\% of field galaxies in the top panels appear to have a boxy bulge, while in the bottom panels this number raises to more than 50\%.} }
    \label{fig:bulgeshapeboxy_hist}
\end{figure}

\begin{figure}
	\includegraphics[width=\columnwidth]{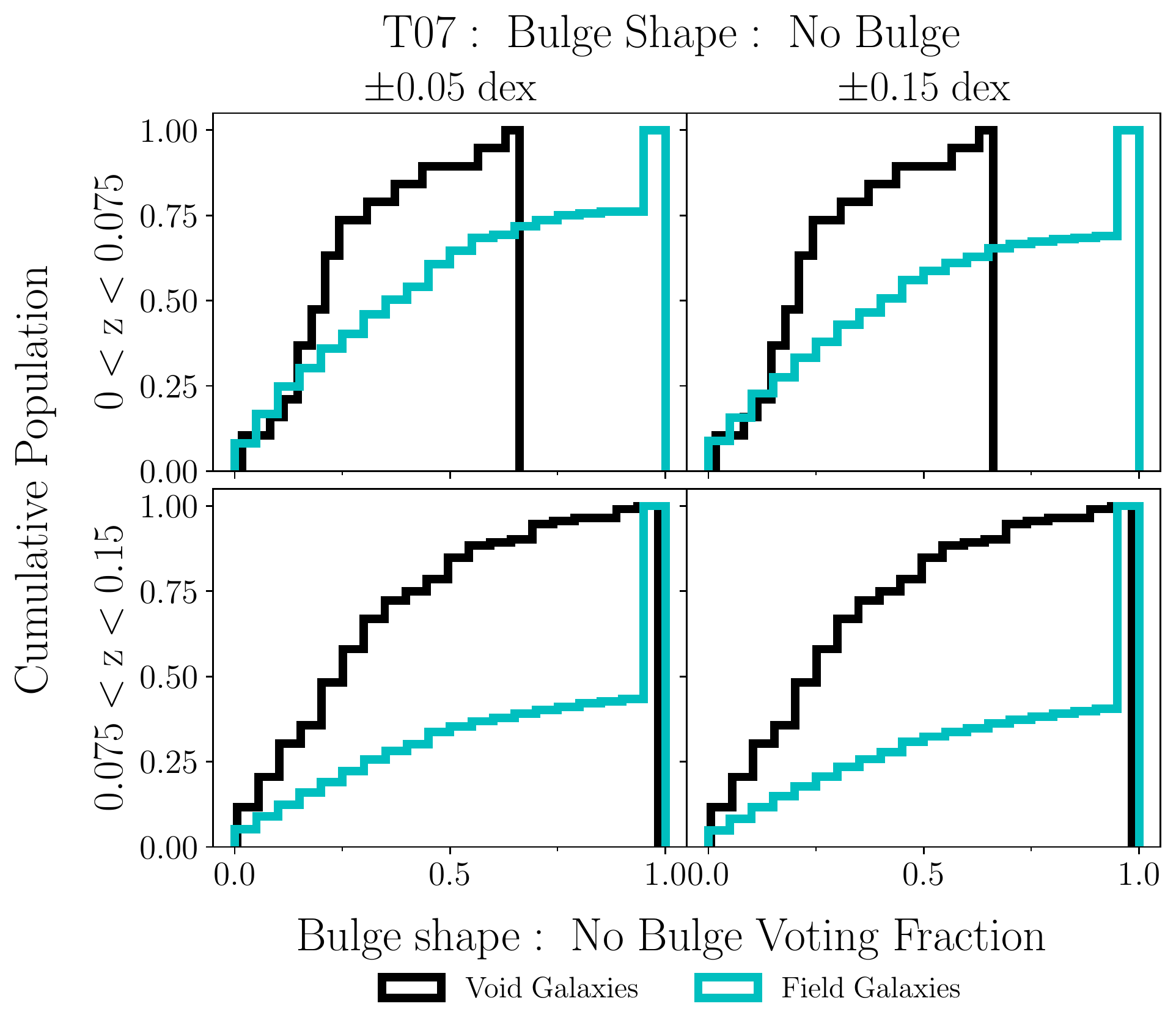}
    \caption{Histogram comparing the voting fraction for question T07 about edge-on galaxies (\rev{"}Does the galaxy have a bulge at its center?\rev{"}) with answers for \revthree{`}no bulge' in both samples, with a normalized frequency. GAMA \revthree{`}twins' are denoted in cyan, and void galaxies are denoted in black. Left panels represent twins chosen within $\pm0.05$ dex of void galaxies, whereas right panels represent twins chosen within $\pm0.15$ dex. \revthree{Similar to Figure~\ref{fig:nobulge_hist}, void galaxies in all panels and field galaxies in the upper panels all appear to have a bulge, while most field galaxies in the lower panels do not.} }
    \label{fig:bulgeshapenobulge_hist}
\end{figure}

\begin{figure}
	\includegraphics[width=\columnwidth]{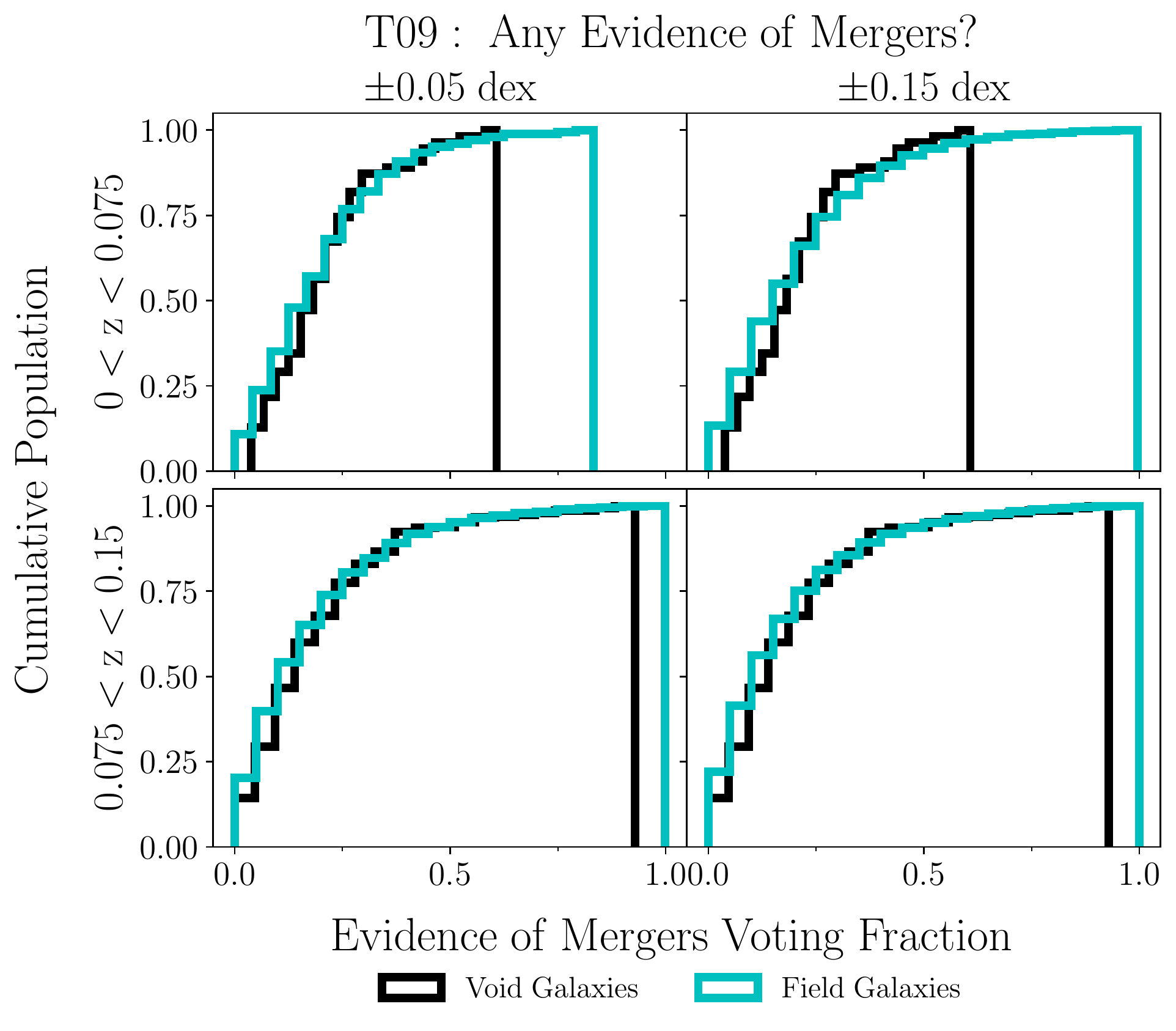}
    \caption{Histogram comparing the voting fraction for question T09 (\rev{"}Is the galaxy currently merging or is there any tidal debris?\rev{"}) with answers for \revthree{`}merging', \revthree{`}tidal debris', or \revthree{`}both' in void and field galaxies, with a normalized frequency. GAMA \revthree{`}twins' are denoted in cyan, and void galaxies are denoted in black. Left panels represent twins chosen within $\pm0.05$ dex of void galaxies, whereas right panels represent twins chosen within $\pm0.15$ dex. \revthree{Mergers do not appear to be occurring in most of the galaxies included in this study.} }
    \label{fig:mergers_hist}
\end{figure}

Next, Figures \ref{fig:bulgeshaperounded_hist}, \ref{fig:bulgeshapeboxy_hist}, and \ref{fig:bulgeshapenobulge_hist} represent the answers for question T07, which asks about the shape of central galaxy bulge (if one exists) for galaxies identified as being viewed edge-on. Test statistics for this question are higher than the face-on group (question T04), suggesting an even higher contrast between the two types of galaxies \revthree{at this viewing angle. All three of these questions appear to be highly significant for all subsamples. Investigating these histograms tells us that people remain in general agreement with edge-on bulges being rounded in all samples (Figure~\ref{fig:bulgeshaperounded_hist}), where Figure~\ref{fig:bulgeshapeboxy_hist} shows a strong disagreement with the presence of boxy bulges in void galaxies.} Similarly, a majority of votes in Figure~\ref{fig:bulgeshapenobulge_hist} shows that people strongly \textit{disagree} that void galaxies have no bulge. Therefore, we can conclude that void galaxies are extremely likely to always have a bulge present when viewed edge-on.

\rev{We \revtwo{again note the} importance of redshift} when it comes to discerning between round and boxy bulges in edge-on galaxies. When viewing a galaxy edge-on, it can be much easier to see that a bulge is present than it is to see the exact shape of said bulge. Especially in less-resolved images, bulges that are actually boxy may appear to be rounded, and we caution against this bias when seeing this result, \rev{and therefore recommend using only the 0 < z < 0.075 sample for forming a conclusion about the shape of an edge-on bulge in question T07.}

Finally, question T09 ("Is the galaxy currently merging or is there any sign of tidal debris?") can be investigated by referencing Figure~\ref{fig:mergers_hist}. For simplicity, we are more concerned with identifying the general presence of mergers as opposed to the identification method (tidal debris, visible mergers, or both). Because question T09 has four possible choices, if we then group the three positive identifications of mergers together into simply, "Evidence of Mergers", it then becomes redundant to include "Neither", as the sum of all four must equal one, and they will then have the same statistics. From this point on, we will only refer to question T09 as, \rev{"Evidence of mergers?" with the general answers being "Yes (presence of tidal debris, visible mergers, and/or both)", or "No". }

This question and all four possible answers are extremely relevant to this study, as the presence of merging galaxies and tidal debris are direct consequences of denser environments. It is with simple logic that we would hypothesize void galaxies have a much lower possibility of either of these occurring due to their isolated nature, but our results contradict this assumption.

\rev{From Figure~\ref{fig:mergers_hist},} there is a strong general disagreement for the presence of mergers in void and comparable field galaxies. We can effectively conclude that there does not appear to be signs of active merging in our sample \rev{,especially because we do not find any form of statistical significance.} However, this Galaxy Zoo question only accounts for mergers in-progress (presence of a merging satellite) or relatively recently (tidal debris). This does not account for past mergers that may be revealed through an analysis of star formation histories. \rev{In addition, because we limit our definition of void galaxy \revthree{`}twins' to be within either $\pm0.05$ or $\pm0.15$ dex in terms of stellar mass and sSFR, this could likely account for the lack of mergers. Merging galaxies are known to cause a significant increase in both star formation rates and stellar mass. Therefore, not allowing for field galaxies to have a significantly higher $\rm M_*$ or sSFR could be why we do not see strong signs of mergers.}

\section{Discussion}
\label{sec:discussion}

Through investigating true void galaxies identified by \citet{Alpaslan2014}, we are able to uncover what (if any) \revtwo{effect the} environment has on local galaxy evolution. For the most part, we find that our results align with previous literature, and any deviations can be logically explained. 

\citet{Rojas2004} uses nearest-neighbor statistics in SDSS to investigate the \rev{S\'ersic} index of void galaxies and \revthree{`}wall' galaxies. Similar to the work done here, \revtwo{they} \rev{employed} the Kolmogorov-Smirnov test to test for \revtwo{significance in the difference between S\'ersic index distributions} of their subsamples (void and wall galaxies, near and far), and found significance in the far sample. \revthree{Here, we also find no K-S significance in our local galaxies,  with it only being evident for our far sample, but with  low test statistics. From Galaxy Zoo, the 0.075 < z < 0.15 sample stands out for a few questions: in many cases, we see extremely high significance that is not replicated in the near sample, 0 < z < 0.075. \revtwo{It is possible that these results are due to the low sample size of our "local" region, as the statistical significance tests (i.e., the Kolmogorov-Smirnov test) are normalized by the sample size, and therefore small sample sizes are less optimal for performing such tests.}} \revtwo{It is also worth notice that the "near" sample from \citet{Rojas2004} uses} a maximum redshift ($\rm z_{max}$) of 0.025, where ours is 0.075, and for their far sample, $\rm z_{max}$ = 0.089, while ours is z = 0.15. \rev{Clearly, how redshift is analyzed has a clear significance to results.}

When it comes to the physical properties, we note that while most previous literature conducts analysis on the star formation rates of void galaxies compared to those in denser environments, we refrain from doing so because we have specifically selected field galaxies to be similar in \rev{specific} star formation (within $\pm0.05$ and $\pm0.15$ dex).  

We remain in general agreement when it comes to previous findings on void galaxy morphology as a whole. We find that void galaxies are dominated by late-type, or disky, galaxies \citep{VanDeWeygaert2011, Beygu2016, Pustilnik2019}. \revthree{We note that we have inconclusive results with regard to findings from \citet{Rojas2005} stating that void galaxies have more spiral galaxies. We find that both field and void galaxies are both dominated by spirals, and Figure~\ref{fig:spiral_hist} appears to show that there are more spirals in void galaxies in the local regime. This is supported by higher test statistics in this redshift range from Table~\ref{table:kstest}. However, the higher redshift field and void galaxies appear to have little difference in the voting fraction of spirals, and show lower test statistics. Nearly all of these test statistics have high significance.} 

\revthree{For question T07 ("Does the galaxy have a bulge at its center?"), we note the difference in samples (and high K-S test statistics), displayed graphically in Figure~\ref{fig:bulgeshapenobulge_hist}.} A large majority of the Galaxy Zoo \revthree{citizen scientists} disagree with the fact that void galaxies have no bulge, indicating that they nearly always have a bulge present, and that this bulge is \revthree{usually} rounded.

From an inside-out galaxy formation perspective, the definite presence of bulges, \rev{particularly those that are obvious/dominant round ones,} makes sense. In such a galaxy formation model, the inner bulge is the oldest part of the galaxy, and \revtwo{slowly accretes} surrounding material to form the disk \citep{Kepner1999, Robertson2004, vanDokkum2010, Nelson2012, Nelson2016}.  In the case of our void galaxies, as a result of their isolation, the lack of material to accrete would result in the bulge being far more dominant than the disk. This follows the results that we see in the Galaxy Zoo voting fractions for obvious and dominant bulges, \rev{the bulge shapes, and the dominance of disk galaxies (which is also supported by the GAMA S\'ersic index).} 

\rev{On the topic of bulges, we found earlier that there is a strong disagreement for the presence of bars and boxy bulges in the samples, especially for the void galaxies (see Figures~\ref{fig:bar_hist} and \ref{fig:bulgeshapeboxy_hist}). There currently exist several arguments in literature, such as those by \citet{Kruk2019} and \citet{Peschken2019}, that bars can be tidally induced. \revtwo{Through} logical reasoning, one could assume that these tidally-induced bars would therefore happen at higher rates in denser environments, which has the potential to explain Figure~\ref{fig:bulgeshapeboxy_hist}. Question T09 in Figure~\ref{fig:mergers_hist} shows little evidence of mergers, one sign of which includes tidal debris, and therefore these two questions may be linked. We previously explained that reducing the accepted stellar mass and specific star formation rates for analogue void galaxies has likely affected our results on mergers. Therefore, if these tidal interactions from nearby galaxies can cause bars to form in galaxies, our reduction in stellar mass and star formation may also be affecting the results for barred (or boxy edge-on) galaxies. }

\section{Conclusions}
\label{sec:conclusions} 

In this paper we presented an overview of void galaxies identified by \citet{Alpaslan2014}, focusing on the properties of \rev{S\'ersic} index, stellar mass, specific star formation rate, and effective radius. In addition, we used the Galaxy Zoo Survey to investigate the morphological voting fractions, with the goal of determining \rev{the typical void galaxy morphology, and} whether void galaxies are morphologically different from their field galaxy counterparts. We can summarize our findings through the following points:

 \begin{enumerate}
     \item \rev{Both void and field galaxies, as seen in the S\'ersic indices (Figure~\ref{fig:sersic_hist}) and presence of features in Galaxy Zoo (Figure~\ref{fig:features_hist}), are dominated by disk galaxies. However, we do not find evidence that void galaxies exhibit a higher fraction of disks.}
     \item In all subsamples of \rev{far} edge-on galaxies, we see strong indicators that the bulges of void galaxies are round as opposed to boxy, \rev{and} results are highly suggestive that void galaxies almost always have a bulge (Figure~\ref{fig:bulgeshapenobulge_hist}, Table~\ref{table:kstest}). \rev{The significant differences in rounded edge-on bulges are also found in our local sample.}
     \item Neither field nor void galaxies appear to show strong evidence of mergers occurring, despite their difference in environment density. \rev{However, this is likely due to our imposed restraint on stellar mass and star formation rates, as mergers are known to cause a strong increase in both quantities.}
     \item We see little difference in the results for how we define the void galaxy counterparts in GAMA (\revthree{`}twins'), \rev{whether we select stellar mass and specific star formation rates within $\pm0.05$ dex or $\pm0.15$ dex,} but redshift appears to have an affect.
 \end{enumerate}
 
 Overall, we see that void galaxies are rather similar to field galaxies, especially in a limited redshift range for the local Universe. However, we do see evidence from our conclusions that point to how isolated galaxies may evolve differently from their counterparts in filaments and tendrils. 
 
 While our results primarily match previous literature, this study still consists of few void galaxies and analogues, therefore relying on a smaller sample size to conduct analyses compared to the wealth of field galaxies available with GAMA. Investigating the star-formation histories of these galaxies, \revtwo{such as using data from studies such as \citet{Bellstedt2020},} would be ideal to determine how these galaxies are fueled, and whether these assembly histories differ for void galaxies. \rev{Similarly, studying the B/D ratios (e.g., \citet{Casura2022}) of these samples could also provide morphological information beyond the scope of this paper.} In addition, employing techniques to identify a larger catalog of void galaxies would be ideal in order to perform further analysis. Future study into the history and morphology of void galaxies would provide a more quantitative understanding of whether they are truly different from those in denser parts of the Universe, and whether their isolation is the specific cause. 

\section*{Acknowledgements}

The material is based upon work supported by National Aeronautics and Space Administration (NASA) Kentucky award no. 80NSSC20M0047 \rev{OGMB 220984} (NASA-REU to B.W. Holwerda and L.E. Porter). 

This research made use of ASTROPY, a community-developed core PYTHON package for astronomy \citep{Astropy2013, Astropy2018}, \revthree{DR4 of the Galaxy and Mass Assembly Survey \citep{Driver2022}, and Galaxy Zoo (Kelvin et al. in prep). We are grateful to members of GAMA and Galaxy Zoo for contributing to this paper.}

\section*{Data Availability}

\revthree{The data for this project are available from the GAMA DR4 website (\url{http://www.gama-survey.org/dr4/}).}









\bsp	
\label{lastpage}
\end{document}